\documentclass[a4paper,12pt]{article}
\usepackage{epsfig}
\usepackage{psfig}
\usepackage{chart}
\begin{document}

\begin{titlepage}

\baselineskip 24pt

\begin{center}

{\Large {\bf Nonabelian Generalization of Electric-Magnetic 
Duality---a Brief Review}}

\vspace{.5cm}

\baselineskip 14pt

{\large CHAN Hong-Mo}\\
chanhm\,@\,v2.rl.ac.uk \\
{\it Rutherford Appleton Laboratory,\\
  Chilton, Didcot, Oxon, OX11 0QX, United Kingdom}\\
\vspace{.2cm}
{\large TSOU Sheung Tsun}\\
tsou\,@\,maths.ox.ac.uk\\
{\it Mathematical Institute, University of Oxford,\\
  24-29 St. Giles', Oxford, OX1 3LB, United Kingdom}

\end{center}

\vspace{.3cm}

\begin{abstract}

A loop space formulation of Yang-Mills theory high-lighting the significance
of monopoles for the existence of gauge potentials is used to derive a 
generalization of electric-magnetic duality to the nonabelian theory.  
The result implies that the gauge symmetry is doubled from $SU(N)$ to 
$SU(N) \times \widetilde{SU}(N)$, while the physical degrees of freedom
remain the same, so that the theory can be described in terms of either 
the usual Yang-Mills potential $A_\mu(x)$ or its dual $\tilde{A}_\mu(x)$.  
Nonabelian `electric' charges appear as sources of $A_\mu$ but as monopoles 
of $\tilde{A}_\mu$, while their `magnetic' counterparts appear as monopoles 
of $A_\mu$ but sources of $\tilde{A}_\mu$.  Although these results have 
been derived only for classical fields, it is shown for the quantum theory
that the Dirac phase factors (or Wilson loops) constructed out of $A_\mu$ 
and $\tilde{A}_\mu$ satisfy the 't~Hooft commutation relations, so that 
his results on confinement apply.  Hence one concludes, in particular, that 
since colour $SU(3)$ is confined then dual colour $\widetilde{SU}(3)$ 
is broken.  Such predictions can lead to many very interesting physical
consequences which are explored in a companion paper.

\end{abstract}

\end{titlepage}

\clearpage

\baselineskip 14pt

The question whether the electric-magnetic duality of electromagnetism
is generalizable to nonabelian Yang-Mills theories is of course a classic 
theoretical problem of fundamental interest in its own right.  Recently,
however, this long-standing question has been given a new urgency by 
the realization that its application to the Standard Model in 
particle physics can lead to an understanding for the existence both
of the Higgs fields required for symmetry breaking and of the three 
generations of fermions experimentally observed, besides offering at the 
same time an explanation for the values of some of the Standard Model's 
many empirical parameters.  

In this paper we briefly review the steps leading to a solution to this
problem suggested a couple of years ago.  We think such a review is 
worthwhile since the material which has been collected over many years 
is scattered widely in the literature.  The problem of duality in gauge 
theories is seen to be intimately related to the existence or otherwise 
of monopoles, which in turn are best described in loop space.  Our present 
review will therefore take us over these few subjects in turn.  We shall 
not cover, however, any of the phenomenological applications of nonabelian 
duality so as to avoid confusing theoretical with practical issues.  
Interested readers are referred to a companion paper \cite{DSMrph98}
for a review of the phenomenological applications.

\setcounter{equation}{0}

\section{Loop Space}

Most of us have learned by experience to work with gauge theory using the 
gauge potentials $A_\mu(x)$ as variables.  But suppose we were to approach 
gauge theory now for the first time, we would probably ask ourselves the 
question whether $A_\mu(x)$ are the right variables to use to describe 
gauge theory.  After all, $A_\mu(x)$ are gauge-dependent and therefore 
physically unobservable.  Would it not be wiser instead to describe a 
physical theory with measurable quantities?

\begin{figure}
\centering
\input{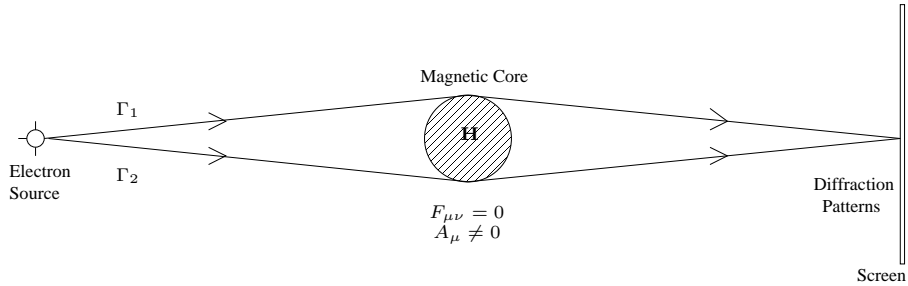}
\caption{Schematic Representation of the Bohm-Aharonov Experiment.}
\label{Bohm}
\end{figure}

Indeed, in classical electrodynamics, the variables used by Faraday and
Maxwell were not the gauge potentials $A_\mu(x)$ but the gauge invariant,
measurable field strengths $F_{\mu\nu}(x)$.  It was only when we started
to deal with quantum mechanics that we were forced to turn to $A_\mu(x)$
as variables.  That this is so is demonstrated by the famous Bohm-Aharonov
experiment \cite{Bohmaha}, as illustrated in Figure \ref{Bohm}.
Although the field strength $F_{\mu\nu}(x)$ vanishes throughout the region
traversed by the charged particle, there are observable effects of the
magnetic field $H$ in the form of diffraction patterns on the screen, showing
that $F_{\mu\nu}(x)$ itself is inadequate to describe completely the 
physical conditions.

The Bohm-Aharonov experiment shows that to describe the quantum mechanics
of a charged particle interacting with an electromagnetic field, the
field strengths $F_{\mu\nu}(x)$ are inadequate and the gauge potentials
$A_\mu(x)$ are sufficient.  But the potentials $A_\mu(x)$ actually give us 
more information than we need.  To describe the diffraction pattern on
the screen, it is already sufficient to know the loop integrals:
\begin{equation}
\alpha_C = e \oint_C A_\mu(x)\, dx^\mu,
\label{alphaC}
\end{equation}
over closed paths $C$, not necessarily the $A_\mu(x)$'s themselves.
Indeed, even $\alpha_C$ are more than necessary, for if all $\alpha_C$
change by integral multiples of $2\pi$, the diffraction pattern will not
be affected.  Thus, what we need are only the phase factors:
\begin{equation}
\Phi_C = \exp ie \oint_C A_\mu(x)\, dx^\mu.
\label{phiC}
\end{equation}
Hence, we conclude, in the words of Wu and Yang, ``The field strength
$F_{\mu\nu}$ underdescribes electromagnetism,  $\ldots$the phase ($\alpha_C$)
overdescribes electro\-mag\-net\-ism$\ldots$.  What provides a complete 
description
that is neither too much nor too little is the phase factor 
($\Phi_C$).''\cite{Wuyang1}  
By `underdescription' here, we mean that different 
physical conditions may correspond to the same values of the variables, 
while by `overdescription', that different values of the variables may 
correspond to the same physical condition.  Hence to have a unique labelling
for the physical conditions in terms of $A_\mu(x)$, one will need to factor
out the classes of $A_\mu(x)$ which are physically equivalent.  In contrast, 
what is nice about the phase factors $\Phi(C)$ is that the same physical
condition corresponds to the same values of $\Phi(C)$, and different
conditions to different values, although any given set of values of 
$\Phi(C)$ need not necessarily correspond to a physical condition.

The situation in nonabelian Yang-Mills theories is similar, except that
here, even in the classical theory, the field strengths $F_{\mu\nu}(x)$
no longer offer a sufficient description \cite{Wuyang2}.  For the quantum 
theory, the gauge potentials $A_\mu(x)$ are again adequate but overdescribe 
the theory, and what provides a complete description for the theory yet 
not an over-description are the path-ordered phase factors (Wilson loops):
\begin{equation}
\Phi_C = P \exp ig \oint_C A_\mu(x)\,dx^\mu.
\label{PhiC}
\end{equation}

Why then do we not use $\Phi_C$ as variables to describe gauge theory?
The reason is that $\Phi_C$ is labelled by the loops $C$ in space-time 
which are infinitely more numerous than the points $x$ in space-time.  
Since the gauge potentials $A_\mu(x)$ labelled by $x$ are already 
sufficient to describe gauge theory, a description in terms of $\Phi_C$
must therefore be highly redundant.  By `redundant' here, we mean that not 
all points in the space spanned by the variables $\Phi(C)$ correspond to
actual physical conditions, but only a subset of it which we may think 
of as a constraint surface in that space.  The `redundancy' being
infinite, the constraint required is bound to be complicated, which 
makes the description in terms of loop variables extremely clumsy.  
Hence, in usual circumstances, one would much rather deal with the 
vagaries of the gauge-dependent $A_\mu$ than with the redundancy of 
$\Phi_C$.  But there are situations some of which we shall discuss, where 
a description in terms of $\Phi_C$ is preferable, indeed may  even be 
necessary.  
In that case we shall need to face the complications and develop the 
formalism for dealing with gauge theory in terms of loop quantities. 

To effect a loop space formulation of gauge theory, our first task would
be to label the loops in space-time, or in other words to introduce some
sort of co-ordinates in loop space.  (It will be seen that it is sufficient 
to consider only those loops passing through some fixed reference point 
$P_0 = \xi_0^\mu$.)  An obvious possibility is to label a loop by the 
space-time co-ordinates of the points on it, thus:
\begin{equation}
C: \ \ \ \{\xi^\mu(s) \colon s = 0 \rightarrow 2\pi,\ \xi(0) = \xi(2\pi)
   = \xi_0 \},
\label{ximus}
\end{equation}
so that $\Phi_C$ in (\ref{phiC}) or (\ref{PhiC}) can be rewritten as:
\begin{equation}
\Phi[\xi] = P_s \exp ig \int_0^{2\pi} ds\,A_\mu(\xi(s)) \dot{\xi}^\mu(s),
\label{Phixi}
\end{equation}
where a dot denotes differentiation with respect to the parameter $s$.
This labelling, however, is again redundant in that if one replaces $s$ by 
another parametrization $s'= f(s)$, it would leave the phase factor 
$\Phi_C$ invariant.  To effect 
a unique labelling of $C$, these reparametrizations should in principle 
be factored out.  Nevertheless, this quotient space is so complicated 
that most people would rather live with the redundancy of the space of 
the loops parametrized by the functions $\xi$.  This 
is the attitude that we shall adopt.  In {\it parametrized loop space}, 
loop quantities such as $\Phi[\xi]$ are just functionals of the
continuous 
(piece-wise smooth)
functions $\xi$ of $s$, which are relatively easy to handle, although care 
has always to be taken in removing the additional redundancy introduced 
by the parametrization.  

Thus, for example, a derivative can be introduced in (parametrized) loop
space just as the functional derivative with respect to $\xi(s)$.  To be 
specific, we shall define the derivative as:
\begin{equation}
\delta_\mu(s) \Psi[\xi] = \lim_{\Delta \rightarrow 0}  \frac{1}{\Delta} 
   \{\Psi[\xi'] - \Psi[\xi]\},
\label{loopderiv}
\end{equation}
with:
\begin{equation}
\xi'^\alpha(s') = \xi^\alpha(s') + \Delta \delta_\mu^\alpha\, \delta(s-s'),
\label{xiprime}
\end{equation}
meaning that the loop is `plucked' by a delta function in the direction $\mu$
at the point on the loop corresponding to the value $s$ of the
parameter.  This definition has to be
interpreted with some care, especially when approximating the continuum by
a discretized space as in lattice theories, where a careless handling may 
easily lead, for example, to asymmetric second derivatives \cite{Corhass,
loop2,Book}. 

Our next task in a loop space formulation is to select the variables for 
describing the gauge field.  In doing so, we have to bear in mind a
major problem already mentioned before in connection with the high degree 
of redundancy in loop variables.  For example, suppose we choose the phase 
factors $\Phi[\xi]$.    If we allow all of these 
$\Phi$'s to take any value in  the gauge group $G$, 
then clearly not all of them will be
expressible in terms of a local gauge potential $A_\mu(x)$ via (\ref{Phixi}), 
there not being enough freedom in $A_\mu(x)$ to satisfy all the conditions 
thereby imposed.  In other words, there are certain sets of values of the 
variables $\Phi[\xi]$ in $G$ which are unphysical.  Thus, in order 
to ensure that in changing to a loop description we are still dealing 
with the same theory though in a different language, we have to impose 
contraints on the values that these loop variables can take so as to 
guarantee that one can recover a local gauge potential $A_\mu(x)$ from 
them.  The ability to write down the appropriate constraints for doing
so is thus one of the first consideration in any loop formulation of gauge
theory.

\begin{figure}
\centering
\input{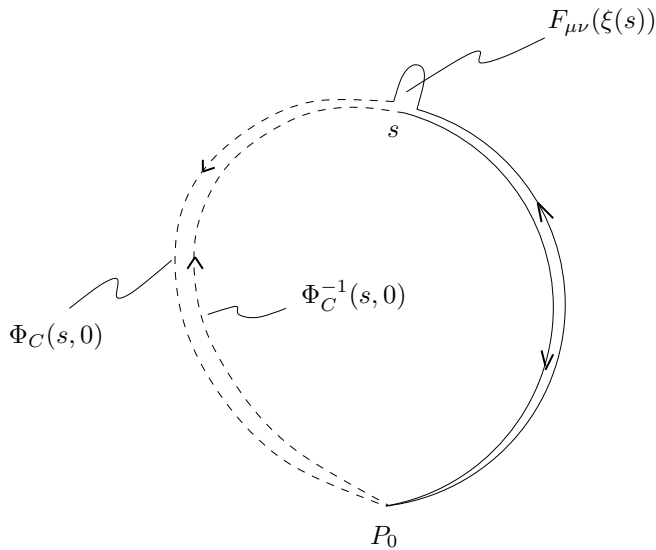}
\caption{Illustration for the quantity $F_\mu[\xi|s]$}
\label{Fmufig}
\end{figure}

For this reason, instead of the seemingly more natural choice of $\Phi[\xi]$
as variables, we choose rather to work with the quantities $F_\mu[\xi|s]$ 
first introduced by Polyakov \cite{Polyakov} as the logarithmic loop 
derivatives of the phase factors $\Phi[\xi]$ in (\ref{Phixi}), namely
\begin{equation}
F_\mu[\xi|s] = \frac{i}{g} \Phi^{-1}[\xi]\, \delta_\mu(s) \Phi[\xi],
\label{Fmuxis}
\end{equation}
Its meaning in space-time is illustrated in Figure \ref{Fmufig},
where it can be seen that $F_\mu[\xi|s]$ depends only on that part of the 
loop before the point labelled by $s$ (hence the special notation $[\xi|s]$ 
for its argument).

The virtue of the quantity $F_\mu[\xi|s]$ lies in the fact that in loop 
space it plays the role of a sort of `connection' similar to that played 
by the gauge potential $A_\mu(x)$ in space-time.  By (\ref{Fmuxis}) it 
tells us how the phase of $\Phi[\xi]$ changes as one moves from one loop 
to a neighbouring loop, i.e. from point to neighbouring point in loop space.  
Of course, when the phase factor $\Phi[\xi]$ exists as a single-valued 
function in loop space, then $F_\mu[\xi|s]$ as given in (\ref{Fmuxis}) 
is trivial as a `connection', corresponding just to what is known in gauge 
theory language as `pure gauge'.  But for $F_\mu[\xi|s]$ as variables
taking arbitrary values the corresponding connection is not in general 
trivial.  Because of this geometrical 
significance, the redundancy-removing contraints take on a particularly 
elegant and physically lucid form, which is intimately related to the
concept of monopoles as topological obstructions in gauge theories.  
The explicit formulation of these constraints has thus to be postponed 
to Section 3 after the concept of monopoles has been introduced.  For 
the moment, we must first prepare some necessary tools.

\begin{figure}
\centering
\input{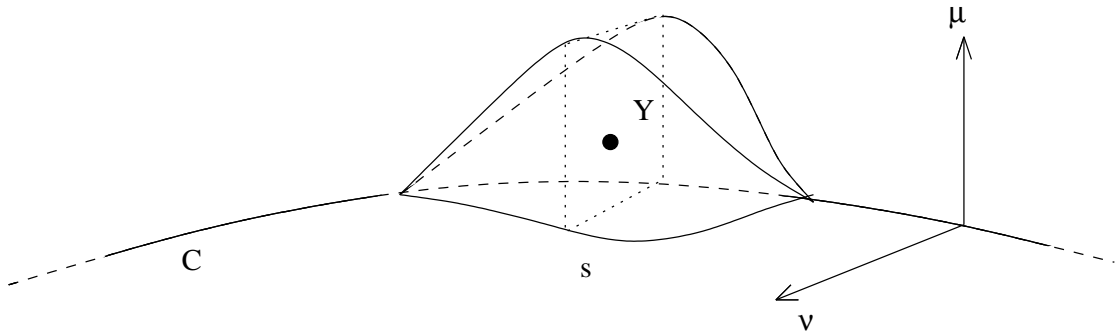}
\caption{Illustration for $G_{\mu\nu}[\xi|s]$.}
\label{skiprope}
\end{figure}

Given the concept of $F_\mu[\xi|s]$ as a `connection' in loop space, one 
can proceed as usual to define a loop space curvature as:
\begin{equation}
G_{\mu\nu}[\xi|s] = \delta_\nu(s) F_\mu[\xi|s]- \delta_\mu(s) F_\nu[\xi|s]
           + ig [F_\mu[\xi|s], F_\nu[\xi|s]].
\label{Gmunu}
\end{equation}
This is the exact parallel of the familiar formula for the field strength 
$F_{\mu\nu}(x)$ in terms of the gauge potential $A_\mu(x)$, and has the 
same geometric significance of a parallel phase transport around an 
infinitesmal circuit, but now in loop space.  Its meaning in space-time,
however, is as illustrated in Figure \ref{skiprope} where the loop `skips' 
over a small 3-volume in space.  For the pure gauge connection in
(\ref{Fmuxis}), the curvature $G_{\mu\nu}[\xi|s]$ is of course zero. 
But in general the curvature need not vanish, in which case, as will be
shown in the next section, it means physically that there is a monopole
charge enclosed inside the small 3-volume `skipped' over by the loop in
Figure \ref{skiprope}.

\begin{figure}
\centering
\includegraphics{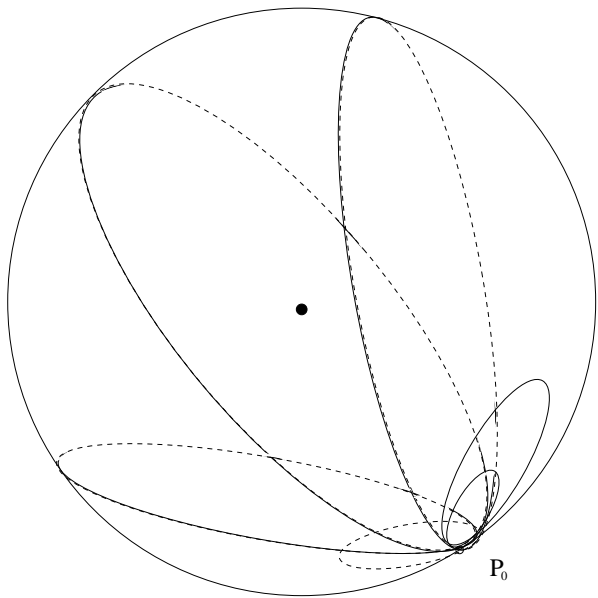}
\caption{Surface swept out by the one-parameter family of loops $\xi_t$.}
\label{looptheloop}
\end{figure}

Further, just as one has constructed from the potential $A_\mu(x)$ the 
phase factor $\Phi[\xi]$ which, as the `holonomy', is an extension
of the concept of curvature to a finite-sized loop, so a holonomy in loop 
space can also be constructed from the `connection' $F_\mu[\xi|s]$ as
\cite{loop1}:
\begin{equation}
\Theta_\Sigma = P_t \exp ig \int_0^{2\pi} dt \int_0^{2\pi} ds\,
   F_\mu[\xi_t|s] \frac{\partial \xi_t^\mu(s)}{\partial t},
\label{Thetasigma}
\end{equation}
where $\Sigma$ denotes the parametrized surface:
\begin{equation}
\Sigma: \{ \xi_t^\mu(s) \colon s =0 \rightarrow 2 \pi, t = 0 
\rightarrow 2 \pi \},
\label{Sigma}
\end{equation}
with:
\begin{equation}
\xi_t^\mu(0) = \xi_t^\mu(2 \pi) = \xi_0^\mu,\ t = 0 \rightarrow 2 \pi,
\label{xit0}
\end{equation}
\begin{equation}
\xi_0^\mu(s) = \xi_{2\pi}^\mu(s) = \xi_0^\mu,\ s = 0 \rightarrow 2 \pi.
\label{xis0}
\end{equation}
The closed surface $\Sigma$ swept out by the one-parameter family of loops
$\xi_t$ is illustrated in Figure \ref{looptheloop}, which may be considered
also as a loop in loop space.  Again, for the `pure gauge' connection
(\ref{Fmuxis}), $\Theta_\Sigma$ is trivial and equals the group identity. 
However, it will not be so when the volume enclosed by $\Sigma$ contains
monopole charges, as we shall see later.

To round off this section, we mention some facts which we shall find 
useful later.  From Figure \ref{Fmufig}, it can be seen that in terms 
of ordinary field variables, $F_\mu[\xi|s]$ is also expressible as 
\cite{Polyakov,Zois}:
\begin{equation}
F_\mu[\xi|s] = \Phi_\xi^{-1}(s,0) F_{\mu\nu}(\xi(s)) \Phi_\xi(s,0)
   \dot{\xi}^\nu(s),
\label{Fmuinx}
\end{equation}
where $\Phi_\xi(s_2,s_1)$ is the parallel phase transport:
\begin{equation}
\Phi_\xi(s_2,s_1) = P_s \exp ig \int_{s_1}^{s_2} ds\, A_\mu(\xi(s)) 
   \dot{\xi}^\mu (s)
\label{Phis2s1}
\end{equation}
from $s_1$ to $s_2$ along the loop $\xi$.  This formula (\ref{Fmuinx}) 
allows us to translate from loop space language back to local field 
language.  For example, by substitution of (\ref{Fmuinx}) into the expression:
\begin{equation}
{\cal A}_F^0 = -\frac{1}{4 \pi \bar{N}} \int \delta \xi \int_0^{2\pi} ds
  \,{\rm Tr}\{F_\mu[\xi|s] F^\mu[\xi|s] \} |\dot{\xi}(s)|^{-2},
\label{CalA0F}
\end{equation}
and performing the functional integral, one obtains the standard pure
Yang-Mills action:
\begin{equation}
-\frac{1}{16 \pi} \int d^4x\, {\rm Tr}\{ F_{\mu\nu}(x) F^{\mu\nu}(x) \}
\label{calA0F}
\end{equation}
in terms of local field variables, if we define the normalization factor
$\bar{N}$ as:
\begin{equation}
\bar{N} = \int_0^{2\pi} ds \int \prod_{s' \neq s} d^4 \xi(s').
\label{Nbar}
\end{equation}
The expression (\ref{CalA0F}) will serve as the loop space field action 
in terms of $F_\mu[\xi|s]$ as variables.

\setcounter{equation}{0}

\section{Monopoles}

Monopoles occur as topological obstructions in gauge theories with 
compact multiply-connected gauge groups.  They may be defined as 
follows \cite{Lubkin,Wuyang1,Coleman}.  Take a 1-parameter family 
of closed loops $\{C_t\}$, as that parametrized by $\xi_t$ in 
(\ref{Sigma}) above, sweeping out a 2-dimensional surface $\Sigma$.
For each $t$ we can then associate a phase factor $\Phi(C_t)$ which is 
an element of the gauge group $G$.  As $t$ varies from $0$ to $2\pi$, 
$\Phi(C_t)$ traces out a closed curve, say $\Gamma_\Sigma$,
in $G$.  If $G$ is multiply-connected, then $\Gamma_\Sigma$ will belong 
to one or other of the homotopy classes $\pi_0(G)$ of closed curves 
in $G$, where members of different classes cannot be continuously 
deformed into one another.  The homotopy class to which $\Gamma_\Sigma$
belongs is defined as the monopole charge enclosed inside the surface
$\Sigma$. 

At first sight, this might seem a rather abstruse definition for a monopole
for which, after all, the primary example is just the magnetic charge of
electromagnetism, which can be represented simply by a novanishing 
divergence of the magnetic field, or in relativistic notation by a nonvanishing
$\partial_\mu {}^*\!F^{\mu\nu}(x)$.  On closer examination, however, it is 
easily seen first, that the above definition reduces in the abelian theory
back to the usual interpretation of the monopole as a source of the field
${}^*\!F$, and secondly, that for a nonabelian theory this latter 
interpretation 
no longer works and cannot be used to define the monopole \cite{Book}.  
Indeed, the above definition, or its equivalent, is the only known valid 
extension of Dirac's magnetic monopole to nonabelian Yang-Mills theory.

This definition exhibits the essentially topological nature of the monopole 
charge which is by definition discrete, and since invariant under continuous 
deformations, also conserved.  The values that this charge can take depend 
on the topological property of the gauge group.  Thus, for (compact) 
electrodynamics, the gauge group is $U(1)$ which has the topology of a 
circle, on which the homotopy classes of closed curves are labelled by
their winding numbers.  As a result, the magnetic charge is quantized, 
meaning that it takes integral values in ${\bf Z}$, as first noted by 
Dirac \cite{Dirac}.  

For simply-connected gauge groups such as $SU(N)$, there can be no monopoles,
there being only one homotopy class of closed curves which contains the 
vacuum.  However, this does not mean that there can be no nonabelian monopoles
in gauge theories with $su(N)$ symmetries.  By an $su(N)$ theory, one usually
means a theory invariant under the gauge Lie algebra $su(N)$.  This by itself
does not specify the gauge group, since different Lie groups can correspond
to the same Lie algebra.  But it is the the global structure of the
gauge group
which determines whether a theory can have monopoles.  Thus, for the pure 
$su(N)$ Yang-Mills theory containing only gauge bosons in the adjoint 
representation and nothing else, the gauge group is $SU(N)/{\bf Z}_N$ 
and not $SU(N)$, since two elements in $SU(N)$ differing by only a factor
$\exp 2i\pi/N$ have the same effect on the gauge boson field and should 
thus be considered as identical elements of the gauge group.  That being
the case, and $SU(N)/{\bf Z}_N$ being $N$-tuply connected, the pure $su(N)$
Yang-Mills theory can have monopoles with charges labelled by elements
of ${\bf Z}_N$.  In particular, pure $su(2)$ Yang-Mills theory has gauge
group $SU(2)/{\bf Z}_2 = SO(3)$ and monopole charges labelled by a sign
$\pm$, with $+$ corresponding to the vacuum and the charge $-$ being 
its own conjugate.  Similarly, pure $su(3)$ Yang-Mills theory has gauge 
group $SU(3)/{\bf Z}_3$ and monopole charges labelled by the cube roots 
of unity $1, \omega, \omega^2$, with $\omega = \exp 2\pi i/3$. 

\begin{figure}
\centering
\includegraphics{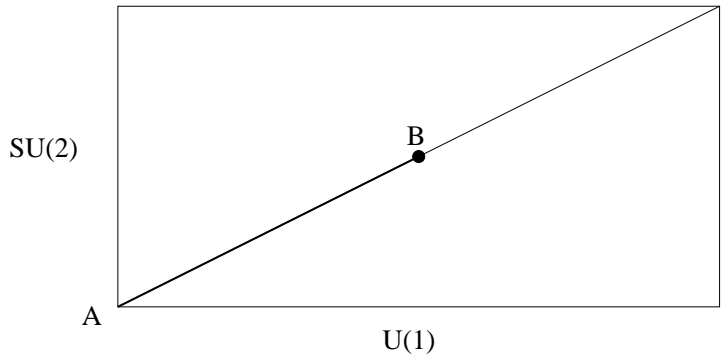}
\caption{Monopoles in the electroweak theory}
\label{U2wind}
\end{figure}

To determine the gauge group and hence whether a theory has monopoles, we
need to examine the gauge transformation properties of all fields present
in the theory \cite{Yang,monocharge}.  Take for example the electroweak 
theory as we know it today which has either:
\begin{itemize}
\item[(i)] $SU(2)$ doublets with half-integral hypercharges, e.g. $(\nu, e)_L$
with $Y = 1/2$; or else:
\item[(ii)] $SU(2)$ singlets or triplets with integral hypercharges, e.g. $e_R$
with $Y = 1$, $A_\mu$ with $Y = 0$.
\end{itemize}
Hence, if we put:
\begin{equation}
\tilde{f} = \exp 2\pi i T_3 \ \ \ \in SU(2)_f,
\label{ftilde}
\end{equation}
and:
\begin{equation}
\tilde{y} = \exp 2\pi i Y   \ \ \ \in U(1)_Y,
\label{ytilde}
\end{equation}
the couple $(f \tilde{f}, y \tilde{y})$ in $SU(2)_f \times U(1)_Y$ has
exactly the same physical effect as $(f, y)$ and therefore has to be 
identified
with the latter.  As a result, the gauge group is not $SU(2) \times U(1)$
but $[SU(2) \times U(1)]/{\bf Z}_2 = U(2)$.  Now, in contrast to $SU(2)$, 
the group $U(2)$ can have monpoles \cite{monocharge}.  
Indeed, as seen in Figure \ref{U2wind}, $AB$ is a closed curve in $U(2)$ 
which cannot be continuously deformed to zero.  It winds half-way round each 
of the $SU(2)_f$ and $U(1)_Y$ subgroups.  Hence a $U(2)$-monopole of unit
charge can be thought of as carrying an $SO(3)$ monopole charge $\eta = -$, 
as well as a $U(1)_Y$ monopole charge of half the Dirac value.  In general,
the monopoles of the electroweak theory are  labelled by an integer
$n$, where a charge $n$ monopole can be thought of as carrying simultaneously:
\begin{eqnarray}
\eta & = & (-1)^n \ \ \ SO(3)\ {\rm monopole\ charge}, \nonumber \\
\tilde{y} & = & \pi n/g_1 \ \ \ U(1)_Y\ {\rm monopole\ charge}.
\label{U2charge}
\end{eqnarray}

A similar analysis carried out for the full Standard Model given the
presently known spectrum of charges reveals that the gauge group is
$[SU(3) \times SU(2) \times U(1)]/ {\bf Z}_6$, and that its monopoles,
also labelled by an integer $n$, can be thought of as carrying 
simultaneously the following charges \cite{monocharge}:
\begin{eqnarray}
\zeta & = & \exp 2\pi i n/3 \ \ \ SU(3)/{\bf Z}_3\ {\rm monopole\ charge}, 
   \nonumber \\
\eta & = & (-1)^n \ \ \ SO(3)\ {\rm monopole\ charge}, \nonumber \\
\tilde{y} & = & 2 \pi n/(3g_1) \ \ \ U(1)_Y\ {\rm monopole\ charge}.
\label{SMcharge}
\end{eqnarray}
This fact will be of use in the physical applications of nonabelian duality
\cite{DSMrph98}.

The lists given in (\ref{U2charge}) and (\ref{SMcharge}) of available 
monopole charges in respectively the electroweak theory and the Standard 
Model are the equivalents of the statement in electromagnetism that the
magnetic charge is quantized.  The well-known Dirac quantization condition
for the abelian theory, however,
contains more information, for it says not only that the magnetic
charge $\tilde{e}$ is quantized but that it is quantized in units of
$1/2e$, which can be thought of as a relation 
\begin{equation}
2 e \tilde{e} = 1
\label{diraccond}
\end{equation}
between 
the minimal coupling strengths.  A parallel for this exists also for 
$su(N)$ theories which for our present normalization convention reads 
as \cite{dualcomm}:
\begin{equation}
g \tilde{g} = 1,
\label{Diraccond}
\end{equation}
and can be similarly derived.\footnote{The couplings here which follow
the original Dirac convention are the so-called unrationalized couplings.  
For the rationalized couplings now more in common use, the conditions 
should read respectively $e \tilde{e} = 2 \pi$ and $g \tilde{g} = 4 \pi$.}

The definition given above for the monopole charge is unfortunately a 
little abstract.  In order to be useful, the monopole charge has 
to be expressed explicitly in terms of whatever field variables one may
choose to adopt.  In terms of the standard variables $A_\mu(x)$, monopole
charges are always a little hard to handle.  This can be seen already in
the abelian theory.  If $A_\mu(x)$ exists and is single-valued, then it
follows that:
\begin{equation}
\partial_\mu F_{\nu\rho} + \partial_\nu F_{\rho\mu} + \partial_\rho F_{\mu\nu}
   = 0,
\label{bianchi}
\end{equation}
or that for:
\begin{equation}
{}^*\!F_{\mu\nu} = -\frac{1}{2} \epsilon_{\mu\nu\rho\sigma} F^{\rho\sigma},
\label{fstar}
\end{equation}
we have:
\begin{equation}
\partial_\mu {}^*\!F^{\mu\nu} = 0.
\label{bianchia}
\end{equation}
However, in the presence of a monopole, $\partial_\mu {}^*\!F^{\mu\nu}$ cannot
vanish.  Hence, $A_\mu(x)$ must be singular somewhere.  This is the reason
for the Dirac string \cite{Dirac}.

The way out, of course, is to consider $A_\mu(x)$ as a patched quantity
\cite{Wuyang1}.  One covers, for example, the sphere by two patches ($N$ 
and $S$):
\begin{eqnarray}
(N): \ \ \ & 0 \leq \theta < \pi, \ \ \ & 0 \leq \phi < 2\pi, \nonumber \\
(S): \ \ \ & 0 < \theta \leq \pi, \ \ \ & 0 \leq \phi < 2\pi,
\label{patchesNS}
\end{eqnarray}
and define $A_\mu^{(N)}$ and $A_\mu^{(S)}$ separately in $N$ and $S$,
with the two potentials related to each other in the overlap region by
a gauge transformation parametrized by the (patching) function:
\begin{equation}
S = \exp ie\alpha,
\label{patchingf}
\end{equation}
where for $S$ to be well-defined the phase $\alpha$ has to change by
an integral multiple of $2\pi$ for $\phi = 0 \rightarrow 2 \pi$, 
giving thus the Dirac quantization condition (\ref{diraccond}).  Similarly 
for nonabelian Yang-Mills theory, one can introduce patched potentials 
to accommodate monopoles, the procedure being then a little more 
complicated.  In either case, however, the patching depends on the locations 
of the monopoles, and the number of patches required increases exponentially 
with the number of monopoles introduced.  It thus appears that a description 
of monopoles in terms of $A_\mu(x)$ is going to be very complicated, so 
much so that even the intrinsic difficulty of loop space formulations 
may now be worth facing by comparison.

The definition above of the monopole charge being given in terms of loop 
quantities in the first place, it is not surprising that it has a simpler
representation in the loop space formulation, especially for the nonabelian 
theory \cite{loop1,Book}.  For illustration, it is sufficient to exhibit 
this explicitly only for the simplest example with gauge group $SO(3)$.  
Recall that $F_\mu[\xi|s]$ is by definition the logarithmic derivative 
of $\Phi[\xi]$.  Hence, we may write:
\begin{equation}
\exp ig dt \int_0^{2\pi} ds\, F_\mu[\xi_t|s] (\partial
   \xi_t^\mu(s)/\partial t)
   \sim \Phi^{-1}[\xi_{t+dt}] \Phi[\xi_t].
\label{increPhi}
\end{equation}
The loop space holonomy $\Theta_\Sigma$ is the product ordered in $t$ 
of such factors and is thus the total change in $\Phi[\xi_t]$ as $t =0
\rightarrow 2\pi$.  Now both $\Phi$ and $\Theta$ may be interpreted as an
element of either the gauge group $SO(3)$ or its double cover $SU(2)$.
However, if we wish to exhibit the monopole charge as an element of
${\bf Z}_2$ considered as a subgroup of $SU(2)$, then we should work
in the latter.  Remembering that the corresponding curve
$\Gamma_\Sigma$ is a closed curve in $SO(3)$, we see that
$\Theta_\Sigma$ must wind around 
$SU(2)$ an odd number of `half-times' if $\Sigma$ contains a monopole charge
$-$, but an even number of `half-times' if $\Sigma$ contains no
monopole.  
Hence we conclude:
\begin{equation}
\Theta_\Sigma = \zeta_\Sigma I,
\label{zetasigma}
\end{equation}
where $\zeta_\Sigma$ is the monopole charge enclosed inside $\Sigma$.
This formula (\ref{zetasigma}) actually holds for any theory with gauge
group $SU(N)/{\bf Z}_N$.

\begin{figure}
\centering
\input{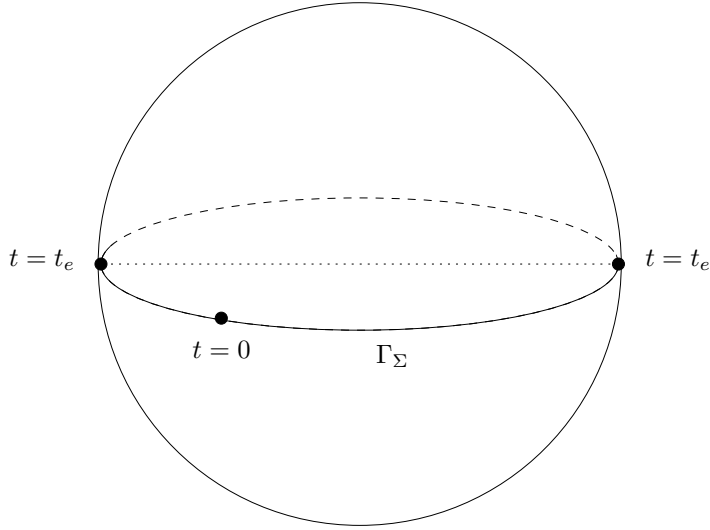}
\caption{A curve representing an $SO(3)$ monopole}
\label{holoncurv}
\end{figure}

It is instructive to examine how this result arises in detail in terms 
of the (patched) gauge potential (Figure \ref{holoncurv}).  
Without loss of generality, we shall
choose the reference point $P_0 = \xi_0^\mu$ to be in the overlap region,
say on the equator which corresponds to the loop $\xi_{t_e}$.  Starting
at $t=0$, where $\Phi^{(N)}[\xi_0]$ is the identity, the phase factor
$\Phi^{(N)}[\xi_t]$ traces out a continuous curve in $SU(2)$ 
(which has the topology of a 3-sphere) until it
reaches $t = t_e$.  At $t = t_e$ one makes a patching transformation
and goes over to $\Phi^{(S)}[\xi_t]$.  From $t = t_e$ onwards, the phase
factor $\Phi^{(S)}[\xi_t]$ again traces out a continuous curve until $t$
reaches $2 \pi$, where it becomes again the identity and joins up with
$\Phi^{(N)}[\xi_0]$.  In order that the curve $\Gamma_\Sigma$ so traced 
out winds only half-way round $SU(2)$ while being a closed curved in 
$SO(3)$, as it should if $\Sigma$ contains a monopole, we must have
\begin{equation}
\Phi^{(N)}[\xi_{t_e}] = - \Phi^{(S)}[\xi_{t_e}],
\label{PhiNSrel}
\end{equation}
which means for the holonomy
\begin{equation}
\Theta_\Sigma = (\Phi^{(S)}[\xi_{t_e}])^{-1}\,\Phi^{(N)}[\xi_{t_e}]
=-I
\end{equation}
as required in (\ref{zetasigma}).

The formula (\ref{zetasigma}) for the monopole charge in terms of the loop
space holonomy $\Theta$, though already explicit, is not as convenient for 
our discussion as the differential formula in terms of the loop space
curvature $G_{\mu\nu}$.  As noted already in Section 1, curvature is 
just an infinitesimal version of the holonomy.  Hence, in the absence of
monopoles $G$ vanishes, but if the loop $\xi$ passes through a monopole
at the point labelled by $s$, then $G_{\mu\nu}[\xi|s]$ must take on a
value equal to the logarithm of the monopole charge $\zeta$ at that point.
Hence, we can write, for a classical monopole moving along the world line
$Y^\mu(\tau)$ \cite{monact1}:
\begin{equation}
G_{\mu\nu}[\xi|s] = - 4 \pi \tilde{g} \int d\tau \kappa[\xi|s] 
   \epsilon_{\mu\nu\rho\sigma} \dot{\xi}^\rho(s) \frac{dY^\sigma(\tau)}{d\tau}
   \delta^4(\xi(s) - Y(\tau)),
\label{Gausslaw}
\end{equation}
where $\kappa[\xi|s]$ satisfies:
\begin{equation}
\exp i \pi \kappa = \zeta.
\label{condkappa}
\end{equation}
This formula has to be interpreted with some care since given $\zeta$
the solution for $\kappa$ in (\ref{condkappa}) in the Lie algebra is not
unique.  But in any case, for $\zeta \neq 1$, $\kappa \neq 0$ which 
means that monopoles can be regarded as sources of curvature in loop
space, as already anticipated.  For the abelian theory, the equation
(\ref{Gausslaw}) reduces to the familiar formula for a classical point
magnetic charge:
\begin{equation}
\partial_\nu {}^*\!F^{\mu\nu}(x) = - 4 \pi \tilde{e} \int d\tau 
\frac{dY^\mu(\tau)}   {d\tau} \delta(x - Y(\tau)).
\label{gausslaw}
\end{equation}

\setcounter{equation}{0}

\section{Wu-Yang Criterion and Poincar\'e Lemma}

An attractive feature of monopoles considered as topological obstructions in 
gauge fields is that their topology defines their own dynamics.  This was 
first pointed out in a beautiful paper by Wu and Yang in 1976 \cite{Wuyangcr}.
That this is so is intuitively clear.  The assertion that there is a monopole
at a certain point $x$ in space-time, as discussed in the last section,
means that the gauge field surrounding $x$ has to have a certain topological
structure, and if the monopole is displaced to another point, then the
gauge field will have to rearrange itself so as to maintain the same
topological structure around the new point.  There is thus naturally a
coupling between the gauge field and the position of the monopole, or in
physical language a topologically induced interaction between the field
and the monopole.

To deduce the explicit form of this interaction, one can proceed as follows.
One writes down first the free action of the gauge field together with that
of a particle.  For example, for electromagnetism and a classical particle
of mass $m$, one has:
\begin{equation}
{\cal A}^0 = {\cal A}^0_F + {\cal A}^0_M,
\label{calA0}
\end{equation}
where ${\cal A}^0_F$ is the usual Maxwell action, and
\begin{equation}
{\cal A}^0_M = - \int d\tau,
\label{calA0M}
\end{equation}
the integral being taken along the world-line of the particle with $\tau$
being the proper time along the world-line.  Extremizing this action with
respect to the dynamical variables of the problem, namely $A_\mu(x)$ for 
the field and co-ordinates $Y^\mu(\tau)$ for the particle, one obtains the 
free equations of the field and of the particle.  Suppose now, however, one 
stipulates that the particle carries a (magnetic) monopole charge and imposes 
on to the system the appropriate topological condition that this should be so.
This condition couples the field and the particle, as already explained, 
so that if one extremizes again the action (\ref{calA0M}) under the
imposed constraint, the equations of motion will no longer be free but
coupled equations involving an `interaction' between the particle and the
field. 

What are the coupled equations so obtained?  Knowing as one does that 
classical electromagnetism is dual symmetric, one is not surprised that
they turn out to be just the dual to the Maxwell and Lorentz equations for 
the motion of an electric charge in an electromagnetic field.  Indeed, this 
was the way that Wu and Yang deduced the equations in their original paper
\cite{Wuyangcr}.  However, a direct attack on the problem so posed is not as 
easy as it might seem at first sight.  The reason is that $A_\mu(x)$, which is
the dynamical variable for the field, is a patched quantity in the presence
of a monopole, as explained in the last section, where the patching depends
on the other dynamical variable $Y^\mu(\tau)$ for the particle.  Extremizing
${\cal A}^0$ with respect to $A_\mu(x)$ is thus not a simple matter, although
(according to Wu in private communication) possible.

There is, however, a very simple and elegant method for solving this 
problem \cite{monact1,Book}.  The trick is to adopt as field variables not
the gauge potential $A_\mu(x)$ but the field strength $F_{\mu\nu}(x)$.  
Being gauge invariant, $F_{\mu\nu}(x)$ is not patch-dependent even in
the presence of a monopole.  Further, the topological condition defining
a (magnetic) monopole charge at $Y^\mu(\tau)$ can be  expressed simply
in
terms of $F_{\mu\nu}(x)$ as (\ref{gausslaw}).  Incorporating (\ref{gausslaw})
as a constraint on the action (\ref{calA0}) by means of a Lagrange multplier 
$\lambda_\mu(x)$ and extremizing with respect to $F_{\mu\nu}(x)$ and 
$Y^\mu(\tau)$, one obtains then easily the equations:
\begin{equation}
F^{\mu\nu}(x) = 4 \pi [\frac{1}{2} \epsilon^{\mu\nu\rho\sigma}
   (\partial_\sigma \lambda_\rho(x) - \partial_\rho \lambda_\sigma(x))],
\label{ELeq1}
\end{equation}
and
\begin{equation}
m \frac{d^2 Y^\mu(\tau)}{d \tau^2} = - 4 \pi \tilde{e} 
   [\partial_\nu \lambda_\mu(Y(\tau)) - \partial_\mu \lambda_\nu(Y(\tau))]
   \frac{d Y^\nu(\tau)}{d \tau}.
\label{ELeq2}
\end{equation}
The equation (\ref{ELeq1}) says that the dual Maxwell field 
${}^*\!F_{\mu\nu}(x)$
is also a gauge field:
\begin{equation}
{}^*\!F_{\mu\nu}(x) = \partial_\nu \tilde{A}_\mu(x) - \partial_\mu 
\tilde{A}_\nu(x),
\label{dfmunu}
\end{equation}
with the (dual) potential:
\begin{equation}
\tilde{A}_\mu(x) = 4 \pi \lambda_\mu(x).
\label{atilde}
\end{equation}
Then using this, one can rewrite the other equation (\ref{ELeq2}) as:
\begin{equation}
m \frac{d^2 Y^\mu(\tau)}{d \tau^2} = - \tilde{e}\, {}^*\!F^{\mu\nu}(Y(\tau))
   \frac{d Y_\nu(\tau)}{d \tau}.
\label{dlorentzeq}
\end{equation}
As expected, these equations, together with the constraint (\ref{gausslaw}), 
are exactly the dual of the equations of motion for an electric charge in an 
electromagnetic field. 

There is an important detail in the above derivation which at first sight
looks like a flaw but which when understood has far reaching consequences 
for the development which follows.  The variables $F_{\mu\nu}(x)$ adopted 
to solve the variational problem are more numerous than the original 
variables $A_\mu(x)$ (there being 6 components to $F_{\mu\nu}$ 
compared with 4 to $A_\mu$)
and must therefore be regarded as `redundant' in the sense the term was
used in Section 1 while discussing loop variables.  In other words, given
a set of values for $F_{\mu\nu}(x)$, there is no guarantee that they can
be derived from a potential $A_\mu(x)$ unless their values are appropriately
constrained.  Now, the beauty of the above derivation is that the dynamical 
constraint (\ref{gausslaw}) imposed, representing the topological definition 
of the monopole charge, already ensures the existence of the potential, 
thereby removing automatically the redundancy in the new variables.  Indeed, 
one notices from (\ref{gausslaw}) that except on the world-line $Y(\tau)$ 
of the monopole, one has:
\begin{equation}
\partial_\nu {}^*\!F^{\mu\nu}(x) = 0,
\label{divfstar}
\end{equation}
which is exactly the condition which implies that $F_{\mu\nu}(x)$ is 
derivable from a potential.  This well-known mathematical fact, to which we 
shall have ample occasions to recall, we shall refer to as the Poincar\'e 
Lemma, although  it is only a very special case of that important 
theorem in differential geometry.  The only place where the condition 
(\ref{gausslaw}) does not guarantee the existence of the potential is on 
the monopole world-line, but that is no problem since at the monopole
position $A_\mu(x)$ should not exist in any case.

The same derivation can be applied also to a quantum particle carrying a
monopole charge.  For example, for a Dirac particle, we replace the action
(\ref{calA0M}) above by \cite{monact2}:
\begin{equation}
{\cal A}^0_M = \int d^4x\,\bar{\tilde{\psi}}(x)\, (i \partial_\mu 
\gamma^\mu - m)\,\tilde{\psi}(x),
\label{calA0Mq}
\end{equation}
and the current on the right-hand side of (\ref{gausslaw}) by its quantum
equivalent:
\begin{equation}
\partial_\nu {}^*\!F^{\mu\nu}(x) = - 4\pi \tilde{e} \bar{\tilde{\psi}}(x) 
   \gamma^\mu \tilde{\psi}(x).
\label{gausslawq}
\end{equation}
Extremizing then the action ${\cal A}^0$ with respect to $F_{\mu\nu}(x)$ 
and $\tilde{\psi}(x)$ under the constraint (\ref{gausslawq}) yields the 
equation (\ref{ELeq1}) together with the equation \cite{monact2}:
\begin{equation}
(i \partial_\mu \gamma^\mu - m) \tilde{\psi}(x) = - \tilde{e} \tilde{A}_\mu(x)
   \gamma^\mu \tilde{\psi}(x).
\label{ELeq2q}
\end{equation}
Again, the equations obtained are exactly the dual for that of an electric
charge as expected.

Our next objective is now to generalize to nonabelian gauge theory to
derive the equations of motion for monopoles.  This is not possible using 
again as variables the field strength $F_{\mu\nu}(x)$ since, in contrast 
to the abelian theory, these are not gauge-invariant and has therefore 
also to be patched in the presence of a monopole.  This difference with 
the abelian theory is of course very deep, and no simple modifications 
are likely to overcome this difficulty.  For this reason, we turn to the 
loop formulation treated in Section 1 where the variables are constructed to be
gauge invariant.\footnote{Actually, as defined, the variables $\Phi[\xi]$ 
and $F_\mu[\xi|s]$ depend on the gauge transformation at the reference
point $P_0 = \xi^\mu_0$, but such a transformation is harmless as far as
patching is concerned.}  We write then the free field action ${\cal A}^0_F$
in terms of the loop variables $F_\mu[\xi|s]$ in the form given in 
(\ref{CalA0F}), and, by the Wu-Yang criterion, impose as a dynamical constraint
the condition that the particle in ${\cal A}^0_M$ should carry a monopole 
charge.  Now, according to (\ref{Gausslaw}), this condition can be written in 
terms of the loop curvature $G_{\mu\nu}[\xi|s]$ as:
\begin{equation}
G_{\mu\nu}[\xi|s] = -4\pi J_{\mu\nu}[\xi|s],
\label{Gausslawj}
\end{equation}
where $J_{\mu\nu}[\xi|s]$ represents the monopole current.  For a 
classical particle, this takes the form \cite{monact1}:
\begin{equation}
J_{\mu\nu}[\xi|s] = \tilde{g}\int d\tau\kappa[\xi|s] 
\epsilon_{\mu\nu\rho\sigma}
   \dot{\xi}^\rho(s) \frac{d Y^\sigma(\tau)}{d\tau} \delta^4(\xi(s)-Y(\tau)),
\label{Jmunu}
\end{equation}
and for a Dirac particle, it takes the form \cite{monact2}:
\begin{equation}
J_{\mu\nu}[\xi|s] = \tilde{g}  \epsilon_{\mu\nu\rho\sigma}
  [\bar{\tilde{\psi}}(\xi(s)) \Omega_\xi(s,0) \gamma^\rho \tau_i 
\Omega_\xi^{-1}(s,0)
  \tilde{\psi}(\xi(s))] \tau^i \dot{\xi}^\sigma(s).
\label{Jmunuq}
\end{equation}
We shall return later to explain the meaning of the operator appearing in
(\ref{Jmunuq}):
\begin{equation}
\Omega_\xi(s,0) =\omega(\xi(s)) \Phi_\xi(s,0).
\label{Omega}
\end{equation}
which will be assigned a rather significant role in the applications of
nonabelian duality to phenomenology.  For the moment, it suffices only
to note that it represents a frame rotation in internal symmetry space.

As it stands, the variational problem posed is straightforward though 
somewhat complicated.  Thus, incorporating the constraint (\ref{gausslawq}) 
into the action by means of the Lagrange multipliers $L_{\mu\nu}[\xi|s]$:
\begin{equation}
{\cal A}' = {\cal A}^0 \int \delta \xi ds\, {\rm Tr} \{ L^{\mu\nu}[\xi|s]
   \{G_{\mu\nu}[\xi|s] + 4\pi J_{\mu\nu}[\xi|s] \}\},
\label{calAp}
\end{equation}
and extremizing with respect to the variables $F_{\mu\nu}[\xi|s]$ and 
$Y^\mu(\tau)$ or $\tilde{\psi}(x)$, one obtains the equations of motion
for the nonabelian monopole.  We give here only the equations for the
more interesting Dirac particle \cite{monact2}, namely:
\begin{equation}
\delta_\mu(s) F^\mu[\xi|s] = 0,
\label{Polyakoveq}
\end{equation}
which, according to Polyakov \cite{Polyakov}, is the loop equivalent of 
the Yang-Mills field equation, and:
\begin{equation}
(i \partial_\mu \gamma^\mu - m) \tilde{\psi}(x) = - \tilde{g} \tilde{A}_\mu(x)
   \gamma^\mu \tilde{\psi}(x),
\label{DiracYMeq}
\end{equation}
where the quantity $\tilde{A}_\mu(x)$ which couples to the Dirac particle
like a dual potential can again be expressed in terms of the Lagrange
multiplier $L_{\mu\nu}[\xi|s]$ but now as a rather complicated functional
integral.

The equations derived for nonabelian monopoles from the Wu-Yang criterion
as outlined above are new since we know as yet of no nonabelian generalization
to the abelian electric-magnetic duality by means of which we were able to
infer in the abelian case the equations for the magnetic charge from those 
of the electric charge.  However, we shall not examine the details of 
these new equations for the present, for we shall be able later to achieve 
a much better appreciation of them.  What is of greater interest at the 
moment is a missing link in the derivation akin to that already encountered 
in the parallel derivation of the abelian equations, namely the question 
of the redundancy of the loop variables $F_\mu[\xi|s]$.  What are the 
necessary constraints on $F_\mu[\xi|s]$ to ensure that one can recover 
from them the original $A_\mu(x)$, and have these constraints been satsified 
in the above treatment?

To answer this question, one has to find a generalization to the nonabelian 
theory of the Poincar\'e Lemma which was used to answer a similar question
in the abelian theory.  Given the large number of loop variables, this at
first sight looks very difficult.  Fortunately, one is able to guess an
answer by following a physical intuition based on what one has learned in 
the abelian case \cite{loop1}.  We recall that by the Poincar\'e Lemma, what 
guaranteed the recovery of the potential at the point $x$ from $F_{\mu\nu}(x)$ 
was the condition (\ref{divfstar}), which means physically that at $x$ there 
is no magnetic charge.  Could it not be then that even for the nonabelian 
theory, the absence of monopole charge at any point $x$ would guarantee 
the local existence of the gauge potential?  If that is true, then our 
derivation is complete, for the dynamical constraint imposed (\ref{Gausslaw}) 
does imply that at all points outside the world-line of the monopole, the 
loop curvature $G_{\mu\nu}$ representing the monopole charge vanishes.

This conjecture is found to be correct.  One is indeed able to derive an 
Extended Poincar\'e Lemma \cite{loop1}, which states that given $F_\mu[\xi|s]$ 
depending on $\xi$ only up to the point $s$ as the notation $[\xi|s]$ in 
the argument implies, and satisfying the transversality condition:
\begin{equation}
F_\mu[\xi|s] \dot{\xi}^\mu(s) = 0,
\label{transvers}
\end{equation}
then a gauge potential $A_\mu(x)$ can be locally constructed in regions 
of space-time where the loop space curvature $G_{\mu\nu}[\xi|s]$ vanishes.  
That there is a transversality condition (\ref{transvers}) on $F_\mu[\xi|s]$, 
is not surprising.  As seen in (\ref{Fmuxis}), the longitudinal component
of $F_\mu[\xi|s]$ represents the logarithmic variation of $\Phi[\xi]$ along
the loop, just as that induced by a reparametrization of the loop.  The 
condition (\ref{transvers}) thus corresponds to the parametrization 
independence of the phase factor $\Phi[\xi]$ and is the price one has to pay 
for working in {\it parametrized loop space} for removing the redundancy
thus introduced, as was explained in Section 1.  In terms of local field
variables, as seen in (\ref{Fmuinx}), it corresponds to the statement that 
$F_{\mu\nu}(x)$ is antisymmetric in its indices $\mu$ and $\nu$.  This 
additional constraint is relatively easy to take care of, affecting little 
the following arguments, and for this reason will largely be ignored.

Apart then from a `proof' of the Extended Poincar\'e Lemma \cite{loop1,Book}, 
which we briefly outline below, the loop space formulation of nonabelian 
Yang-Mills theory introduced in Section 1, and the extension of the Wu-Yang 
criterion to the nonabelian Yang-Mills case, are now both complete.

We shall indicate how one can arrive at an Extended Poincar\'e Lemma only
for the case when there are no monopole charges anywhere in space. This 
will be sufficient to illustrate the idea.  For the more general case
with a number of isolated monopole charges, which is of interest to the 
above application of the Wu-Yang criterion, some more technical sophistication 
is required \cite{loop1} but the idea remains similar.

\begin{figure}
\centering
\includegraphics{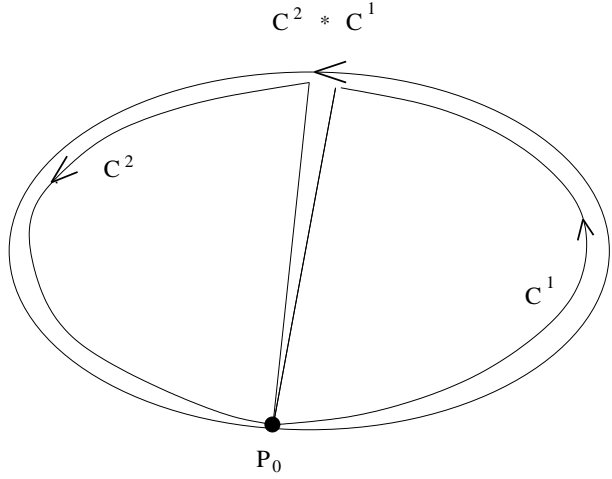}
\caption{Illustration for the composition law for loops}
\label{complawf}
\end{figure}

\begin{figure}
\centering
\input{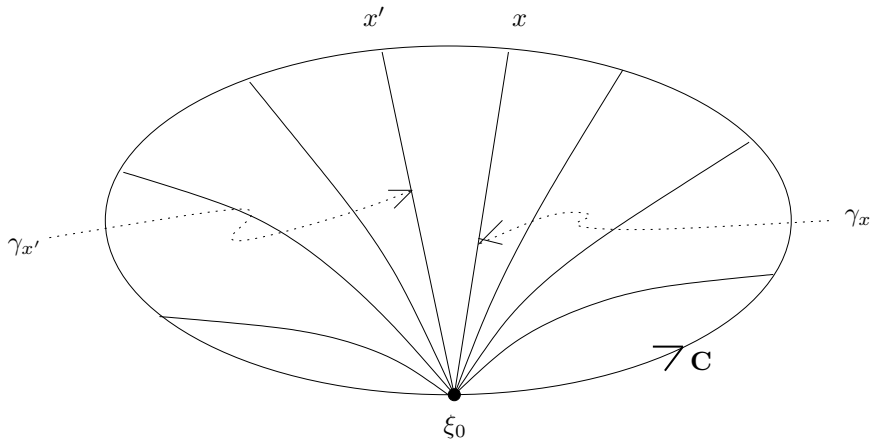}
\caption{Construction of $\Phi(C)$ from $A_\mu(x)$}
\label{fanshape}
\end{figure}

First, we note that if all $G_{\mu\nu}[\xi|s]$ vanish, then $\Theta_\Sigma$
as defined in (\ref{Thetasigma}) equals the identity for all parametrized 
surfaces $\Sigma$.  Recalling that $\Sigma$ is a loop in loop space, this 
implies that the following integral over an open path in loop space:
\begin{equation}
\Theta_\Sigma(t,0) = P_{t'} \exp ig \int_0^t dt' \int_0^{2\pi} ds
   F_\mu[\xi_{t'}|s] \frac{\partial \xi_{t'}^\mu(s)}{\partial t'},
\label{Theta0t}
\end{equation}
is path-independent and is a function only of the `end-point' $\xi_t$.
We define then the phase factor as:
\begin{equation}
\Phi[\xi_t] = \Theta_\Sigma^{-1}(t,0),
\label{Phiintheta}
\end{equation}
which by definition gives $F_\mu[\xi|s]$ as its logarithmic loop derivative.
Second, by construction, one shows, from the fact that $F_\mu[\xi|s]$ is
transverse and depends on $\xi$ only up to the point $s$, that $\Phi[\xi]$
obeys the composition law, namely that:
\begin{equation}
\Phi(C^2*C^1) = \Phi(C^2) \Phi(C^1),
\label{complaw}
\end{equation}
where $C$ denotes the geometric loop in space-time corresponding to the
parametrized loop $\xi$, and $C^2*C^1$ represents the loop obtained by 
first going around $C^1$ then $C^2$, as illustrated in Figure \ref{complawf}.
Third, to every point $x$ in space-time, draw a straight line $\gamma_x$
joining the reference point $P_0 = \xi^\mu_0$ to $x$ and construct the
phase factor $\Phi$ for the triangle formed by $\gamma_{x'}^{-1}*
\gamma_{x'x}*\gamma_x$:
\begin{equation}
h(x',x) = \Phi(\gamma_{x'}^{-1}*\gamma_{x'x}*\gamma_x),
\label{Phitriang}
\end{equation}
where $\gamma_{x'x}$ is the straight line joining $x$ to a neighbouring 
point $x'$.  Define then the gauge potential $A_\mu(x)$ as:
\begin{equation}
A_\mu(x) = -\frac{i}{g} \lim_{\Delta \rightarrow 0} \{h(x',x) - 1\}
\label{Amudef}
\end{equation}
for
\begin{equation}
x'^\nu = x^\nu + \Delta \delta_\mu^\nu.
\label{xprime}
\end{equation}
Using the composition law as illustrated in Figure \ref{fanshape}, one 
shows that any $\Phi(C)$ can indeed be written in terms of this $A_\mu(x)$
in the usual manner.  The recovery of the gauge potential from the
loop variables $F_\mu[\xi|s]$ is then complete.

The above construction of $A_\mu(x)$ depends on the choice of the straight
line $\gamma_x$ for each space time point $x$.  This choice is not unique
and can be replaced by any other choice of an open path for $\gamma_x$
linking the reference point to $x$ so long as neighbouring points are
assigned neighbouring paths.  A different choice for the paths $\gamma_x$,
however, is seen to correspond just to a gauge transformation on $A_\mu(x)$
\cite{loop1,Book}.

\setcounter{equation}{0}

\section{Electric-Magnetic Duality and its Non\-abel\-ian Generalization}

In vacuo, the field tensor $F_{\mu\nu}$ in Maxwell theory satisfies the
equations:
\begin{equation}
\partial_\nu F^{\mu\nu}(x) = 0,
\label{maxwell}
\end{equation}
and:
\begin{equation}
\partial_\nu {}^*\!F^{\mu\nu}(x) = 0,
\label{maxwellstar}
\end{equation}
which are symmetric under the *-operation interchanging electricity
with magnetism.  This property of the theory is what is loosely known 
as electric-magnetic duality.  In particular, just as the equation 
(\ref{maxwellstar}) guarantees that the field $F_{\mu\nu}$ is derivable
from a potential $A_\mu(x)$,
\begin{equation}
F_{\mu\nu}(x) = \partial_\nu A_\mu(x) - \partial_\mu A_\nu(x),
\label{fina}
\end{equation}
so also the equation (\ref{maxwell}) implies that there is a local `dual' 
potential $\tilde{A}_\mu(x)$ such that:
\begin{equation}
{}^*\!F_{\mu\nu}(x) = \partial_\nu \tilde{A}_\mu(x) - \partial_\mu 
\tilde{A}_\nu(x).
\label{fstarinat}
\end{equation}
And just as $F_{\mu\nu}$ is invariant under the gauge transformation:
\begin{equation}
A_\mu(x) \longrightarrow A_\mu(x) + \partial_\mu \alpha(x),
\label{atransf}
\end{equation}
for an arbitrary $\alpha(x)$, so is also ${}^*\!F_{\mu\nu}$ invariant under
the gauge transformation:
\begin{equation}
\tilde{A}_\mu(x) \longrightarrow \tilde{A}_\mu(x) + \partial_\mu 
   \tilde{\alpha}(x),
\label{attransf}
\end{equation}
for another arbitrary $\tilde{\alpha}(x)$ which need have nothing to do with
the $\alpha(x)$ in (\ref{atransf}).  Hence, it follows that the theory 
is itself invariant under a doubled $U(1) \times \tilde{U}(1)$ gauge 
transformation where, apart from having the opposite parity (because of
the *), the `magnetic' $\tilde{U}(1)$ as a group is the same as the `electric'
$U(1)$.  It is important to note, however, that although there are two 
independent gauge degrees of freedom as represented by (\ref{atransf}) and 
(\ref{attransf}), the physical degrees of freedom remain  that given 
by either $F_{\mu\nu}$ or ${}^*\!F_{\mu\nu}$, but not both, since these two 
quantities are always related by the algebraic relation (\ref{fstar})
defining the Hodge star *.  The logical steps in the above arguments 
are summarized in Chart \ref{chart1}.

What happens when there are charges around?  The Maxwell theory as usually
formulated is then not dual symmetric because there are only electric but
no magnetic charges.  This is, however, merely a whim of nature; the 
theory itself still keeps the dual symmetry for there is in prinicple
nothing (apart from experiment) to stop one introducing magnetic charges
into the theory and write for the equations of motion:
\begin{equation}
\partial^\nu F_{\mu\nu}(x) = j_\mu(x),
\label{maxwelle}
\end{equation}
and:
\begin{equation}
\partial^\nu {}^*\!F_{\mu\nu}(x) = \tilde{\jmath}_\mu{x},
\label{maxwellm}
\end{equation}
with $j_\mu$ and $\tilde{\jmath}_\mu$ as respectively the electric and magnetic
currents.  In the standard description in terms of the Maxwell field
$F_{\mu\nu}$ and $A_\mu$, electric charges appear as sources of the
field as per {(\ref{maxwelle}), while magnetic charges, as Dirac has
taught us and as we have explained in Section 2, will appear as monopoles 
or topological obstructions in $A_\mu$.  However, if one chooses to
describe Maxwell theory in terms of the dual fields ${}^*\!F_{\mu\nu}$ and
$\tilde{A}_\mu$, magnetic charges instead will appear as sources as per
(\ref{maxwellm}) while electric charges will appear as monopoles.  We
can thus apply the Wu-Yang criterion to either electric or magnetic
charges to derive their equations of motion.  In particular, one sees
that the standard Lorentz and Dirac equations for electric charges can
be deduced as consequences of their topology when regarded as monopoles.
The logical structure of duality for Maxwell theory with charges is
summarized in Chart \ref{chart2}I.

The question now is whether the above derivation is generalizable to 
nonabelian Yang-Mills theory.  It is easy to see that a naive extension
with the same *-operation as dual transform would fail.  Although the 
field tensor satisfies in vacuo:
\begin{equation}
D^\nu F_{\mu\nu}(x) = 0,
\label{Yangm}
\end{equation}
and, if $F_{\mu\nu}$ is derivable from a potential $A_\mu$, also the
Bianchi identity:
\begin{equation}
D^\nu {}^*\!F_{\mu\nu}(x) = 0,
\label{Yangmstar}
\end{equation}
these two equations, despite appearances, are not dual symmetric.  The 
covariant derivative $D_\mu$ occurring in both (\ref{Yangm}) and 
(\ref{Yangmstar}) involves the potential $A_\mu$ of $F_{\mu\nu}$, whereas
for the second equation to be dual to the first, it ought instead to 
involve a potential, say $\tilde{A}_\mu$, bearing the same relation to
the dual field ${}^*\!F_{\mu\nu}$ as $A_\mu$ to $F_{\mu\nu}$, namely:
\begin{equation}
{}^*\!F_{\mu\nu}(x) = \partial_\nu \tilde{A}_\mu(x) - \partial_\mu 
\tilde{A}_\nu(x)
   + i \tilde{g}\,[\tilde{A}_\mu(x), \tilde{A}_\nu(x)],
\label{FstarinAt}
\end{equation}
and this $\tilde{A}(x)$ has no reason to be the same as $A_\mu$.  Indeed, 
one does not even know whether such an $\tilde{A}_\mu$ exists.  In contrast 
to the abelian theory where the equation of motion, namely the Maxwell 
equation (\ref{maxwell}), implies by the Poincar\'e Lemma that a potential 
exists for ${}^*\!F_{\mu\nu}$, the same is not true for the the 
nonabelian case.  
The equation of motion, which is here the Yang-Mills equation (\ref{Yangm}),
does not imply a potential for the dual field ${}^*\!F_{\mu\nu}$.  Worse in
fact, since Gu and Yang \cite{Guyang} have exhibited many explicit 
counter-examples of solutions to the equation (\ref{Yangm}) for which 
no potential 
for ${}^*\!F$ exists.

Does that mean then that electric-magnetic duality is not generalizable 
to Yang-Mills theory?  Not necessarily, for it can be that if one defines 
the dual transform in a different way from *, then duality is retrieved.  
Supposing this to be true, let us first try to imagine what sort of 
properties such a generalized dual transform will need to possess.  We
would want, of course, the new transform to reduce to the Hodge star for 
the abelian theory as a special case, and to be reversible apart for a 
sign, like the Hodge star: ${}^*\!({}^*\!F) = - F$, so as to qualify 
as a 'duality'.  
But the most difficult property to satisfy is that discussed in the 
preceding paragraph of the existence of a dual potential $\tilde{A}_\mu$ 
and in order to recover that, we turn back to the abelian theory for 
inspiration.  The reason why abelian duality works with the Hodge star 
lies in the fact that the source-free Maxwell equation (\ref{maxwell}) 
implies that there are no monoples in the dual field ${}^*\!F$, which is in 
physical terms exactly the condition required by the Poincar\'e Lemma 
for the existence of a gauge potential.  Thus it appears that a crucial 
property of the new transform we seek is that the dual field should be
so defined as to make the source-free Yang-Mills equation (\ref{Yangm}) 
equivalent to the statement that there are no monopole charges in the 
dual field.  Given that the occurrence or otherwise of monopole charges 
in a nonabelian field was shown in our previous sections to be best 
expressed in loop space language, it is indicated that the same language 
be adopted also for constructing the generalized dual transform.

These, then, are the leads which set us off to look for a generalized dual 
transform in loop space \cite{dualsymm}.  To avoid obscuring the basic 
arguments with details, we shall begin with an outline of the construction 
and only return later to consider the `proof' of its validity as given 
in \cite{dualsymm}.   

First, in place of the $F_\mu[\xi|s]$ that we have employed up to now 
as variables in loop space, we introduce a new set: 
\begin{equation}
E_\mu[\xi|s] = \Phi_\xi(s,0) F_\mu[\xi|s] \Phi_\xi^{-1}(s,0),
\label{Emuxis}
\end{equation}
with $\Phi_\xi(s,0)$ defined as in (\ref{Phis2s1}).  These $E_\mu[\xi|s]$
are not gauge invariant like $F_\mu[\xi|s]$ and may not be as useful in general
but seem more convenient for dealing with duality.  In particular, the 
condition that there be no monopole charge anywhere in space which, 
according to Section 3, is expressible in terms of $F_\mu[\xi|s]$ as the 
vanishing of the loop curvature $G_{\mu\nu}[\xi|s]$, is given in terms
of $E_\mu[\xi|s]$ simply as:
\begin{equation}
\delta_\nu(s) E_\mu[\xi|s] - \delta_\mu(s) E_\nu[\xi|s] = 0.
\label{curlE}
\end{equation}
On the other hand, the source-free condition (\ref{Yangm}) for Yang-Mills 
fields which was expressed by Polyakov \cite{Polyakov} as the vanishing of
the loop divergence of $F_\mu[\xi|s]$ (\ref{Polyakoveq}) remains simply:
\begin{equation}
\delta^\mu(s) E_\mu[\xi|s] = 0
\label{divE}
\end{equation}
in terms of $E_\mu[\xi|s]$.

Using these new variables $E_\mu[\xi|s]$ for the field, we now define their 
`dual' $\tilde{E}_\mu[\eta|t]$ as:
\begin{eqnarray}
& & \omega^{-1}(\eta(t)) \tilde{E}_\mu[\eta|t] \omega(\eta(t)) \\
&=& -\frac{2}{\bar{N}} \epsilon_{\mu\nu\rho\sigma} \dot{\eta}^\nu(t)
     \int\!\delta \xi ds\, E^\rho[\xi|s] \dot{\xi}^\sigma(s) \dot{\xi}^{-2}(s)
     \delta(\xi(s) -\eta(t)),
\label{Dualtransf}
\end{eqnarray}
where $\omega(x)$ is a (local) rotation matrix tranforming from the frame in 
which the orientation in internal symmetry space of the fields $E_\mu[\xi|s]$ 
are measured to the frame in which the dual fields $\tilde{E}_\nu[\eta|t]$ 
are measured.  The dual transform \ $\tilde{\ }$\ is by construction
reversible apart 
from a sign, meaning that $\tilde{\tilde{E}} = - E$, and also reduces to 
the Hodge star for the abelian theory, as required.  Further, it can be 
shown by differentiating (\ref{Dualtransf}) that:
\begin{eqnarray}
\lefteqn{
 \omega^{-1}(\eta(t)) \{\delta_\nu(t) \tilde{E}_\mu[\eta|t] - \delta_\mu(t)
   \tilde{E}_\nu[\eta|t] \} \omega(\eta(t)) =} \nonumber\\
&&\!\!\!\!\!\!\!\!\! - \frac{1}{\bar{N}}\!\!\int\!\!
   \delta \xi ds \epsilon_{\mu\nu\rho\sigma}  \{ \dot{\eta}^\beta(\!t\!)
   \dot{\xi}^\alpha(\!s\!)\!-\!\dot{\eta}^\alpha(\!t\!) 
   \dot{\xi}^\beta(\!s\!) \}
   \delta_\rho(\!s\!) E^\rho[\xi|s] \dot{\xi}^{-2}(\!s\!) 
   \delta(\xi(\!s\!)\!-\!\eta(\!t\!)).
\label{curlEtdivE}
\end{eqnarray}
Now equation (\ref{curlEtdivE}) means that so long as the Yang-Mills field 
is source-free and therefore (\ref{divE}) is satisfied, then
\begin{equation}
\delta_\nu(t) \tilde{E}_\mu[\eta|t] - \delta_\mu(t) \tilde{E}_\nu[\eta|t] = 0,
\label{curlEt}
\end{equation}
or that there are no monopoles in the dual field.  Hence, by the Extended 
Poincar\'e Lemma discussed in Section 3, we deduce that a dual potential
$\tilde{A}_\mu(x)$ will exist in that case. 

With this generalized dual transform in hand, we can now analyse the 
dual structure of nonabelian Yang-Mills theory.  We shall begin with
the pure theory with no source either in $E$ or $\tilde{E}$, for which
the structure is summarized in Chart \ref{chart6}.  Consider first the
left half.
The field action is ${\cal A}^0$ which can be expressed either in terms
of the gauge potential $A_\mu(x)$ via the field tensor $F_{\mu\nu}(x)$, 
or in terms of the loop variable $E_\mu[\xi|s]$.  Adopting $A_\mu(x)$
as variables, the theory can be developed along conventional lines.
Extremizing ${\cal A}^0$ with respect to $A_\mu(x)$, one obtains the
standard Yang-Mills equation (\ref{Yangm}), which can be written
also in terms of $E_\mu[\xi|s]$ as (\ref{divE}).  By (\ref{curlEtdivE}),
however, this is equivalent to (\ref{curlEt}) and implies the existence
of a dual gauge potential $\tilde{A}_\mu(x)$.  This chain of arguments 
is summarized in the left-most column of Chart \ref{chart6}.  Alternatively,
adopting $E_\mu[\xi|s]$ as variables which are redundant, the constraint
(\ref{curlE}) has to be imposed, which is incorporated via Lagrange
multipliers $W_{\mu\nu}[\xi|s]$ into the action ${\cal A}$.  Extremizing
now ${\cal A}$ with respect to $E_\mu[\xi|s]$, one obtains an equation
relating $E_\mu[\xi|s]$
to the Lagrange multipliers, from which relation the dual gauge
potential $\tilde{A}_\mu(x)$ can then be constructed.  This chain of
arguments is summarized in the second column of Chart \ref{chart6}.  
Furthermore,
the existence of the dual potential under gauge transformation of which 
the theory is invariant,  together with the original invariance 
under gauge transformations of $A_\mu(x)$, implies
that the theory has overall a 
doubled gauge symmetry $SU(N) \times \widetilde{SU}(N)$.

The dual tranform (\ref{Dualtransf}) allows one to express the field action
${\cal A}^0$ also in terms of the dual loop variables $\tilde{E}_\mu[\xi|s]$.
The fact now that the dual transform is reversible means that in going
over into the dual representation, the steps outlined in the preceding
paragraph can all be repeated with only occasional changes in signs as 
shown on the right half of Chart \ref{chart6}.  Comparing this Chart
with Chart \ref{chart1} 
shows a very close analogy with the abelian theory.

Next, we turn to Yang-Mills theory with charges.  In this case, one works
exclusively with loop variables which admit an easier implementation of the 
Wu-Yang criterion.  Consider first again the left-half of Chart
\ref{chart7}.  The 
free field-particle system as represented by the free action ${\cal A}^0$ 
is subjected to the constraint:
\begin{equation}
\delta_\nu(s) E_\mu[\xi|s] - \delta_\mu(s) E_\nu[\xi|s] 
   = - 4\pi J_{\mu\nu}[\xi|s],
\label{Jmunuconst}
\end{equation}
which is both the constraint imposed by the Wu-Yang criterion for 
deriving the dynamics of the monopoles and that required by the Extended 
Poincar\'e Lemma of Section 3 for removing the redundancy in the loop
variables.  Whether one is dealing with a classical or a Dirac particle
is distinguished by the choice of the free particle action and the form
of the current $J_{\mu\nu}[\xi|s]$, as shown in the separate columns of
Chart \ref{chart7}.  Incorporating the constraint (\ref{Jmunuconst}) 
into the action
${\cal A}$, and extremizing with respect to the field variables $E_\mu[\xi|s]$
and particle variables $Y^\mu(\tau)$ or $\psi(x)$, one obtains the equations
of motion.  One notes in particular the equation for the Dirac monopole,
which couples the monopole to the dual potential $\tilde{A}(x)$ and takes
a form exactly dual of that for a `colour' charge.  That the monopole is
so coupled to the field confirms that the dual potential constructed does
play the expected role of a connection giving parallel phase transport
for the monopole wave function.  It also implies that the theory has an
$\widetilde{SU}(N)$ symmetry corresponding to the phase of monopole wave
functions in addition to the original $SU(N)$ gauge symmetry corresponding 
to the phase of the wave functions of `colour' charges, giving the theory
thus in all an $SU(N) \times \widetilde{SU}(N)$ gauge symmetry.  The equation 
for the classical monopole is perhaps less familiar but is also exactly dual 
to the so-called Wong equation \cite{Wong} for a classical `colour' charge.  
Again, given that the dual transform is reversible, one obtains a near 
symmetry apart from signs between the left and right halves of the Chart.

This then completes our outline of nonabelian duality barring the derivations
of some formulae that we have used, which, unfortunately, will involve some 
rather delicate operations in loop space, and are as yet far from
rigorous.  Even 
more than before, the lack of a general loop calculus, already mentioned 
in Section 1, is here strongly felt.  We shall not go through all the details
which the reader can find in our original publications \cite{dualsymm,dualsym},
but shall just pick up a few representative points for illustration.

First, let us return to the variables $E_\mu[\xi|s]$ defined in (\ref{Emuxis}).
Recalling the definition (\ref{Fmuxis}) of the Polyakov variable $F_\mu[\xi|s]$
one sees that $E_\mu[\xi|s]$ can be pictured as the bold curve in Figure 
\ref{Emuxisfig} where the phase factors $\Phi_\xi(s,0)$ in (\ref{Emuxis}) 
have cancelled parts of the faint curve representing $F_\mu[\xi|s]$.  In 
contrast to $F_\mu[\xi|s]$, therefore, $E_\mu[\xi|s]$ depends really only 
on a ``segment'' of the loop $\xi$ from $s_-$ to $s_+$.  Notice that the
\begin{figure}
\centering
\input{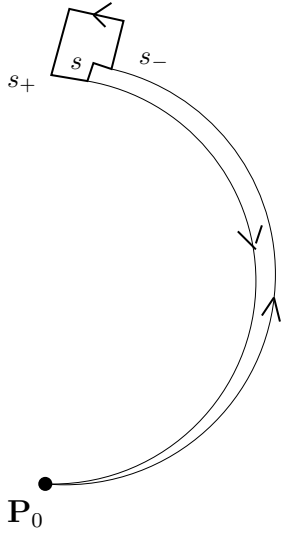}
\caption{Illustration for $E_\mu[\xi|s]$}
\label{Emuxisfig}
\end{figure}
$\delta$-function $\delta(s-s')$ inherent in our definition (\ref{loopderiv})
and (\ref{xiprime}) of the loop derivative $\delta_\mu(s)$ is represented in 
the figure as a bump function centred at $s$ with width $\epsilon = s_+ - s_-$.
The reason for doing so is that, as usual in most functional formulations, 
our treatment here involves operations with the $\delta$-function which 
need to be ``regularized'' to be given a meaning.  Our procedure is to 
take first the $\delta$-function as a bump function with finite width
$\epsilon$ and height $h$, and afterwards take the zero width limit.  In
(\ref{loopderiv}) and (\ref{xiprime}), the limit $\epsilon \rightarrow 0$ 
with $\Delta = \epsilon h$ held fixed is taken first, to be followed by 
the limit $h \rightarrow 0$.  

For example, suppose we wish to take the loop derivative $\delta_\nu(s)$ of the
quantity $E_\mu[\xi|s]$ at the same value of $s$.  Clearly, a loop derivative
has a meaning only if there is a segment of the loop on which it can operate.  
Therefore, to define this derivative, we shall first regard $E_\mu[\xi|s]$ 
as a segmental quantity dependent on the segment of the loop $\xi$ from 
$s - \epsilon/2$ to $s + \epsilon/2$.  We then define the loop derivative 
$\delta_\nu(s)$ using the normal procedure on this segment, and afterwards 
take the limit $\epsilon \rightarrow 0$.  Following this procedure, we have
then by (\ref{Emuxis}) and (\ref{Fmuxis}):
\begin{equation}
\delta_\nu(s') E_\mu[\xi|s] = \Phi_\xi(s,0) \{ \delta_\nu(s') F_\mu[\xi|s]
   + ig \theta(s-s') [F_\nu[\xi|s'], F_\mu[\xi|s]] \} \Phi_\xi^{-1}(s,0),
\label{difEmuxis}
\end{equation}
where $\theta(s)$ is the Heaviside $\theta$-function, so that:
\begin{equation}
G_{\mu\nu}[\xi|s] = \Phi_\xi^{-1}(s,0) \{\delta_\nu(s) E_\mu[\xi|s]
   - \delta_\mu(s) E_\nu[\xi|s]\} \Phi_\xi(s,0).
\label{GmunuinE}
\end{equation}
This shows that the absence of monopoles, which was given in terms of 
the variables $F_\mu[\xi|s]$ by the vanishing of the loop curvature 
$G_{\mu\nu}[\xi|s]$ is here translated in terms of $E_\mu[\xi|s]$ to read 
as the vanishing of the `curl' of $E_\mu[\xi|s]$ as was claimed in 
(\ref{curlE}) above.

Next, as another example, let us check that our generalized dual transform
(\ref{Dualtransf}) does reduce to the Hodge star when the theory is abelian.
To see this, we let the segmental width of $\tilde{E}_\mu[\eta|t]$ in
(\ref{Dualtransf}) go to zero so that we can use the formula given in 
(\ref{Fmuinx}) for $F_\mu[\xi|s]$ to write the left-hand side in terms of 
local quantities:
\begin{equation}
\omega^{-1}(x) \tilde{F}_{\mu\nu}(x) \omega(x) = -\frac{2}{\bar{N}} 
   \epsilon_{\mu\nu\rho\sigma} \int\! \delta\xi ds \,E^\rho[\xi|s] 
   \dot{\xi}^\sigma(s) \dot{\xi}^{-2}(s) \delta(x - \xi(s)).
\label{reducedual1}
\end{equation}
To evaluate the right-hand side, we recall that our procedure is to do 
the integral before taking the width of the segment in $E_\mu[\xi|s]$ to 
zero.  In other words, within the integral, the loop $\xi$ can still vary by 
a $\delta$-functional bump as illustrated in Figure \ref{reducedualfig} (a).  
\begin{figure}
\centering
\input{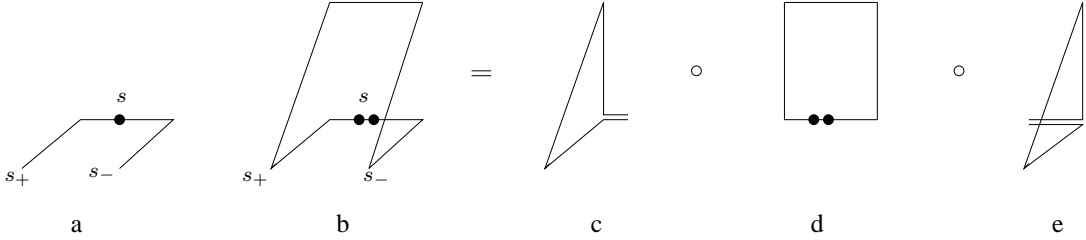}
\caption{Illustration for the Integrand in Dual Transform}
\label{reducedualfig}
\end{figure}
 For such a $\xi$, $E_\mu[\xi|s]$, which is obtained by making a 
$\delta$-functional variation along the direction $\mu$, will take on 
the shape depicted in Figure \ref{reducedualfig} (b).  This last figure 
can be expressed as the product of three factors, namely Figures
\ref{reducedualfig}(c),(d),(e) in the order indicated.  In the abelian
theory, the ordering of the factors is unimportant so that the factors of
Figures (c) and (e) cancel in the limit when the segmental width 
$\epsilon \rightarrow 0$, leaving only the factor of Figure (d), which can 
as before be expressed as $F_{\mu\alpha}(\xi(s)) \dot{\xi}^\alpha(s)$, 
giving:
\begin{eqnarray}
\tilde{F}_{\mu\nu}(x) & = & -\frac{2}{\bar{N}} \epsilon_{\mu\nu\rho\sigma} 
   \int\! \delta\xi ds \,F^{\rho\alpha}(\xi(s)) \dot{\xi}_\alpha(s) 
   \dot{\xi}^\sigma(s) \dot{\xi}^{-2}(s) \delta(x-\xi(s)) \nonumber \\
   & = & - \mbox{\small $\frac{1}{2}$} \epsilon_{\mu\nu\rho\sigma} 
   F^{\rho\sigma}(x).
\label{reducedual}
\end{eqnarray}
This is just the Hodge star relation if we identify $\tilde{F}_{\mu\nu}(x)$
with $\mbox{\mbox{}$^*\!$} F_{\mu\nu}(x)$.  On the other hand, for a 
nonabelian theory, the factors of Figures \ref{reducedualfig} (c) and (e) 
cannot be commuted through the factor of Figure (d) so that the above 
reduction to the Hodge star relation will not go through.  This last 
statement is important for otherwise our claim of nonabelian duality 
under the generalized dual transform (\ref{Dualtransf}) would be in 
contradiction with the result of Gu and Yang quoted above \cite{Guyang}.

These two examples, we hope, should give a taste of the sort of arguments 
one had to go through to justify the results claimed above.  They serve
to illustrate the level of rigour one is able at present to achieve, which 
is unfortunately not as high as one could wish.  Except for this reservation, 
one could claim that a nonabelian generalization of electric-magnetic 
duality is now established.   

\setcounter{equation}{0}

\section{'t~Hooft's Order-Disorder Parameters}

As in electromagnetism, the dual symmetry in nonabelian Yang-Mills theory 
is established only for classical fields.  In contrast to electromagnetism, 
however, where the classical theory has already a wide range of applications, 
Yang-Mills theory applied to physics involves almost always the quantum 
theory.  The construction of a quantized version of the above results 
looks difficult and only the most tentative of beginnings of an attempt 
have so far been made \cite{rudiments}.  However, one very useful result 
for the quantum theory has already been derived which is the subject 
of this section.

In his famous study of the confinement problem in nonabelian gauge 
theories, 't~Hooft \cite{thooft} introduced 2 loop-dependent operators 
$A(C)$ and $B(C)$ which satisfy the following commutation relation:
\begin{equation}
A(C) B(C') = B(C') A(C) \exp(2\pi i l/N)
\label{ABcomm}
\end{equation}
for $su(N)$ symmetry and any 2 spatial loops  $C$ and $C'$ with linking 
number $l$ between them.  $A(C)$ is given explicitly as:
\begin{equation}
A(C) = {\rm Tr} \left[P \exp ig \oint_C A_i(x) dx^i \right],
\label{AC}
\end{equation}
and in the words of 't~Hooft measures the magnetic flux through $C$
while creating electric flux along $C$.   On the other hand, $B(C)$
measures the electric flux through $C$ while creating magnetic flux
along $C$, and plays thus an exactly dual role to $A(C)$.  For lack of
a dual potential, however, $B(C)$ was not given a similar explicit
expression to (\ref{AC}).

Now, in the preceding section, one claims that there does indeed exist 
a dual potential $\tilde{A}_\mu$ for nonabelian gauge fields, although 
by duality is now meant no longer the Hodge star but a more complicated 
transform which reduces to the Hodge star only for the abelian theory.  
That being the case, one ought to have:
\begin{equation}
B(C) = {\rm Tr} \left[ P \exp i \tilde{g} \oint_C \tilde{A}_i(x) dx^i \right]
\label{BC}
\end{equation}
as the explicit expression for $B(C)$, with the dual coupling $\tilde{g}$
related to $g$ by a Dirac quantization condition.  In other words, starting
from the formulae (\ref{AC}) and (\ref{BC}) with $A_\mu$ nd $\tilde{A}_\mu$
related in the manner detailed in the preceding section, one ought to be
able to deduce the commutation relation (\ref{ABcomm}) required by `t~Hooft.
The success in doing so would be an important check on the consistency of
the proposed framework for duality.  It also means that the duality as
defined above accords with what 't~Hooft called duality so that the 
important results he derived are applicable to the present case.

That (\ref{ABcomm}) is indeed satisfied for (\ref{AC}) and (\ref{BC}) is
shown in \cite{dualcomm}.  To appreciate easier how this obtains, recall
first how the parallel assertion to (\ref{ABcomm}) can be deduced in the
abelian case:
\begin{equation}
\left[ ie \oint_C A_i(x) dx^i,\; i \tilde{e} \oint_{C'} \tilde{A}_i(x) dx^i
   \right] = 2 \pi l.
\label{abcomm}
\end{equation}
Using Stokes' theorem, the second integral over $C'$ in (\ref{abcomm})
can be written as a surface integral, thus:
\begin{equation}
- i \tilde{e} \int\!\int_{\Sigma_{C'}} {}^*\!F_{ij}\, d\sigma^{ij},
\label{intfstar}
\end{equation}
or, by the definition of the Hodge star, in terms of the electric field
strength ${\cal E}_i = F_{0i}$ as:
\begin{equation}
i \tilde{e} \int\!\int_{\Sigma_{C'}} {\cal E}_i\, d\sigma^i,
\label{intcalE}
\end{equation}
where $\Sigma_{C'}$ is some surface both spanning and bounded by $C'$.

Consider first the simple case for linking number 1 between $C$ and $C'$.
The loop $C$ in that case will intersect the surface $\Sigma_{C'}$ at some 
point $x_0$.  ($C$ may of course intersect $\Sigma_{C'}$ more than once, 
but the extra intersections occurring pairwise with opposite orientations, 
their contributions to the commutator will all cancel, leaving in effect
just one intersection.)  Except at this point $x_0$, all points on $C$ 
are spatially separated from points on $\Sigma_{C'}$ so that, using the
canonical commutation relation between $A_i(x)$ and ${\cal E}_i(x)$:
\begin{equation}
[{\cal E}_i(x), A_j(x')] = i\, \delta_{ij}\, \delta(x-x')
\label{eacomm}
\end{equation}
we have:
\begin{equation}
\left[ ie \oint_C A_i(x) dx^i, i \tilde{e} \int \!\! \int_{\Sigma_{C'}}
   {\cal E}_j d \sigma^j \right] =i e \tilde{e},
\label{expcomm}
\end{equation}
which by the Dirac quantization condition (\ref{diraccond})(see footnote
for the rationalized couplings adopted here) gives the answer (\ref{abcomm}) 
for $l = 1$ as required.  In case $C'$ winds around $C$ more than once, 
say $l$ times, then $C$ will intersect $\Sigma_{C'}$ at effectively $l$ 
points for each of which the above applies, so that (\ref{abcomm}) still 
remains valid.

What happens when we generalize to the nonabelian case?  Then $A(C)$
and $B(C)$ are each a trace of an ordered product of noncommuting factors, 
{\it not}, in spite of appearances (due to the somewhat misleading standard
notation), an exponential of a line integral for which Stokes' Theorem 
applies, so that the above arguments no longer work.  Nevertheless, one 
finds that one may still associate with each a surface in an analogous 
fashion \cite{dualcomm}.  Take $B(C')$, for example.  The phase factor:
\begin{equation}
\tilde{\Phi}(C') = P \exp i \tilde{g} \oint_{C'} \tilde{A}_i dx^i
\label{PhiCp}
\end{equation}
of which $B(C')$ is the trace, can be written, according to \cite{dualsymm}, 
as:
\begin{equation}
\tilde{\Phi}(C') = \prod_{t = 0 \rightarrow 2 \pi} (1 - i \tilde{g} 
   \tilde{W}[\eta|t]) \sim \prod_{t = 0 \rightarrow 2 \pi}
   \exp (- i \tilde{g} \tilde{W}[\eta|t])\,,
\label{PhiCpp}
\end{equation}
where $\eta$, for $t = 0 \rightarrow 2 \pi$, is a parametrization of
$C'$, and $\tilde{W}$ is a segmental quantity the `loop gradient' of
which, $\delta_\mu \tilde{W}$, as indicated in Chart \ref{chart6}, is the field
variable $\tilde{E}_\mu$.  One can thus write symbolically:
\begin{equation}
\tilde{W}[\eta|t] = \int_{\eta_0}^{\eta(t)} \delta \eta^{' \nu}(t)
   \,\tilde{E}_\nu[\eta'|t],
\label{Wtildeint}
\end{equation}
where the integral in (\ref{Wtildeint}) denotes a `segmental' integral
along some path from a reference point $\eta_0$ to the point $\eta(t)$,
so that in ordinary space, this path appears as a ribbon.  Piecing 
such ribbons together as $\eta(t)$ moves along $C'$ in (\ref{PhiCpp}),
one obtains a surface $\Sigma_{C'}$ spanning over and bounded by $C'$
as suggested.  In as much as the reference point $\eta_0$ and the path
joining it to $\eta(t)$ are both arbitrary for (\ref{Wtildeint}) to
hold, one can choose $\Sigma_{C'}$ to be completely space-like.  This 
surface will again intersect the loop $C$ of $A(C)$ at some point $x_0$.  
(Previous remarks in the abelian case about multiple intersections and 
higher linking numbers between $C$ and $C'$ will still apply and need
not be repeated.) 

The remaining arguments for deriving (\ref{ABcomm}) \cite{dualcomm} then 
follow along much the same lines as for (\ref{abcomm}) in the abelian case 
although they are naturally more complicated because of the noncommutative 
quantities involved.  Thus, starting from the formula (\ref{Wtildeint}), we 
write $\tilde{E}[\eta|t]$ via the dual transform (\ref{Dualtransf}) given in 
the last section in terms of its dual $E[\xi|s]$, which is then related to
the usual Yang-Mills local field tensor $F_{\mu\nu}(x)$.  The commutation
with the operator $A(C)$ using the canonical commutation relation between
$A_i^\alpha(x)$ and the `electric' field strengths ${\cal E}_i^\beta(x)
= F_{0i}^\beta$ then picks out the one point of intersection $x_0$ between
the loop $C$ and the surface associated to $B(C')$ as described in the last 
paragraph.  Again, the Dirac quantization condition comes in in relating
the dual charge $\tilde{g}$ to $g$ giving rise to the factor $2\pi$ in
the exponent on the right-hand side of (\ref{ABcomm}).  For details, the
reader is referred to the original reference \cite{dualcomm}.

Although the derivation of (\ref{ABcomm}) represents but a small step on
the road to quantizing the classical formalism developed in the previous
sections, it is of much practical importance in allowing one to apply 
't~Hooft's far-reaching results \cite{thooft}, which have been used to 
good effect in the Dualized Standard Model \cite{DSMrph98}.  Furthermore, 
as a by-product, one has obtained in (\ref{BC}) an explicit formula for 
the operator $B(C)$ which may be useful in the problem of confinement.  
Although other, and presumably equivalent, explicit formulae for $B(C)$ 
have previously been suggested \cite{Zeni}, they are, in contrast to
(\ref{BC}), usually given as dependent not only on the loop $C$ but also 
on the particular surface spanning it.

\setcounter{equation}{0}

\section{Concluding Remarks}

Starting from a loop space formulation of Yang-Mills theory which helps 
to clarify the role that monopoles play in defining the dynamics and in
guaranteeing the existence of gauge potentials, a generalization of
electric-magnetic duality to the nonabelian theory is derived.  The
result implies in particular that the gauge symmetry is doubled, say
from $SU(N)$ to $SU(N) \times \widetilde{SU}(N)$, where the second
factor has opposite parity to the first.  However, the physical degrees 
of freedom, of course, remain the same, and the theory can be described
in terms of either the usual Yang-Mills potential $A_\mu(x)$ or a dual
potential $\tilde{A}_\mu(x)$.  The theory then admits both nonabelian 
`electric' and nonabelian `magnetic' charges, where the former appear 
as sources of $A_\mu$ but as monopoles of $\tilde{A}_\mu$, while the 
latter appear as monopoles of $A_\mu$ but sources of $\tilde{A}_\mu$.

Although these results have been derived only for classical fields, one
result is known for the quantum theory, namely that the Dirac phase factors
(or Wilson loops) constructed out of $A_\mu$ and $\tilde{A}_\mu$ satisfy
the 't~Hooft commutation relations, so that his results apply.  Hence one 
concludes, in particular, that if $SU(N)$ is in the confined phase then 
its dual $\widetilde{SU}(N)$ is in the Higgs phase, and vice versa.

When applied to the Standard Model with symmetry $su(3) \times su(2)
\times u(1)$, one concludes first that each symmetry has its dual and
particles can in principle carry an `electric' as well as a `magnetic' charge
of each symmetry.  Secondly, using the corollary stated above to `t~Hooft's
result, one concludes that since colour $su(3)$ is confined, dual colour
$\widetilde{su}(3)$ is broken, and since electroweak $su(2)$ is in the
Higgs phase, dual electroweak $\widetilde{su}(2)$ is in the confined phase.  
If we assume that these effects have correspondence in Nature, then their
manifestations should lead to very interesting consequences.  A possible 
scenario for the physical realization of these effects is the subject of our
companion paper \cite{DSMrph98}.

\vspace{1cm}

\noindent {\large \bf Acknowledgement}

\vspace{.3cm}

It is a pleasure to recall our most enjoyable collaborations at different 
periods with respectively Peter Scharbach and Jacqueline Faridani from 
which most of the material reviewed in this paper is drawn.  The present
version is based on some lectures given to the lattice gauge theory group
at Pisa at the invitation of Adriano Di Giacomo.  We are much indebted to
him and to Angela Buonocorso and Kenichi Konishi for their kind hospitality
during this visit.

\clearpage

\begin{chart}
\vspace*{-2cm}
  \centerline{
    \resizebox{!}{23cm}
{\includegraphics{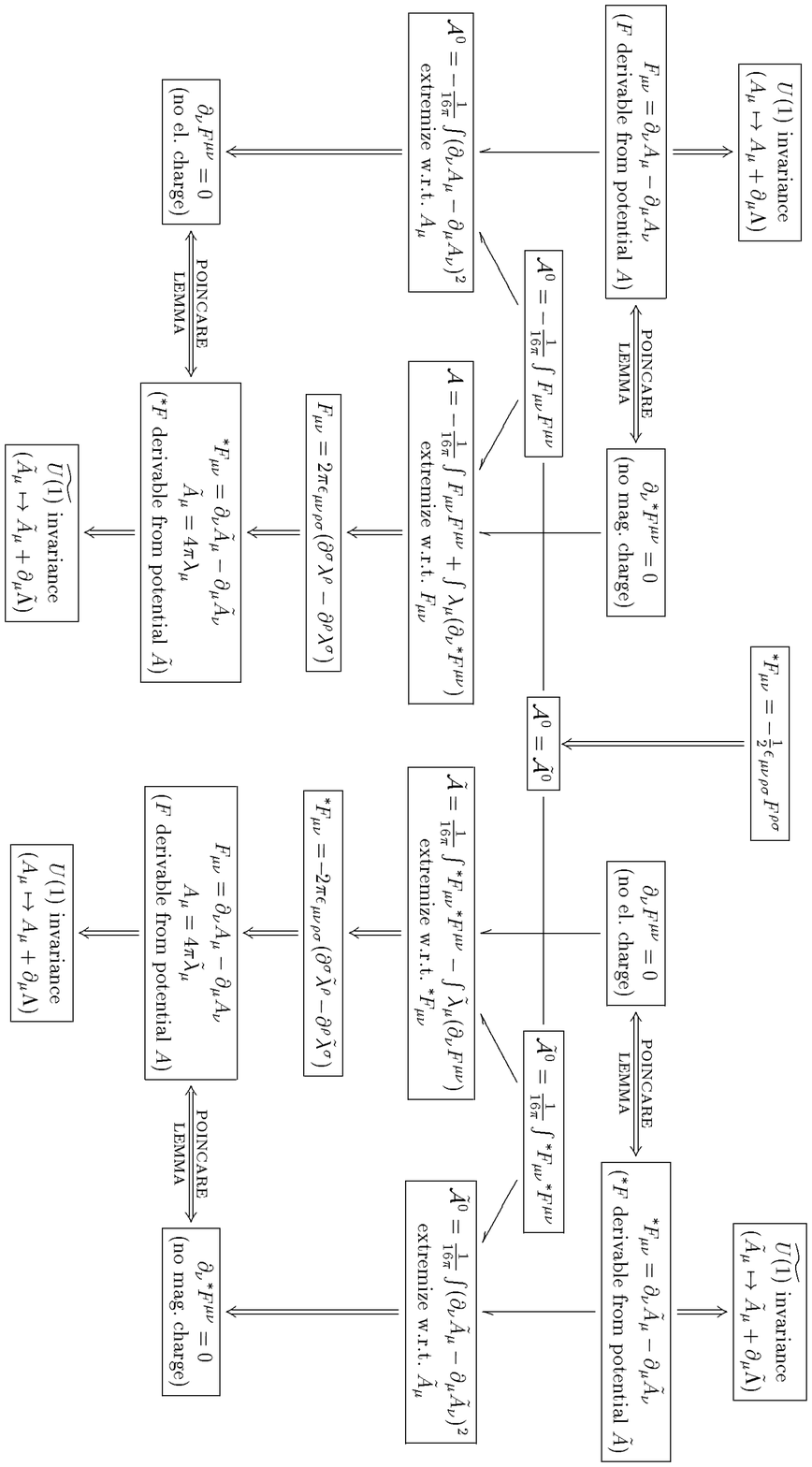}}
    }
\vspace*{-1.5cm}
 \caption[]{Pure Electromagnetism.}
  \label{chart1}
\end{chart}

\clearpage

\begin{chart}
\vspace*{-2cm}
  \centerline{
    \resizebox{!}{23cm}
{\includegraphics{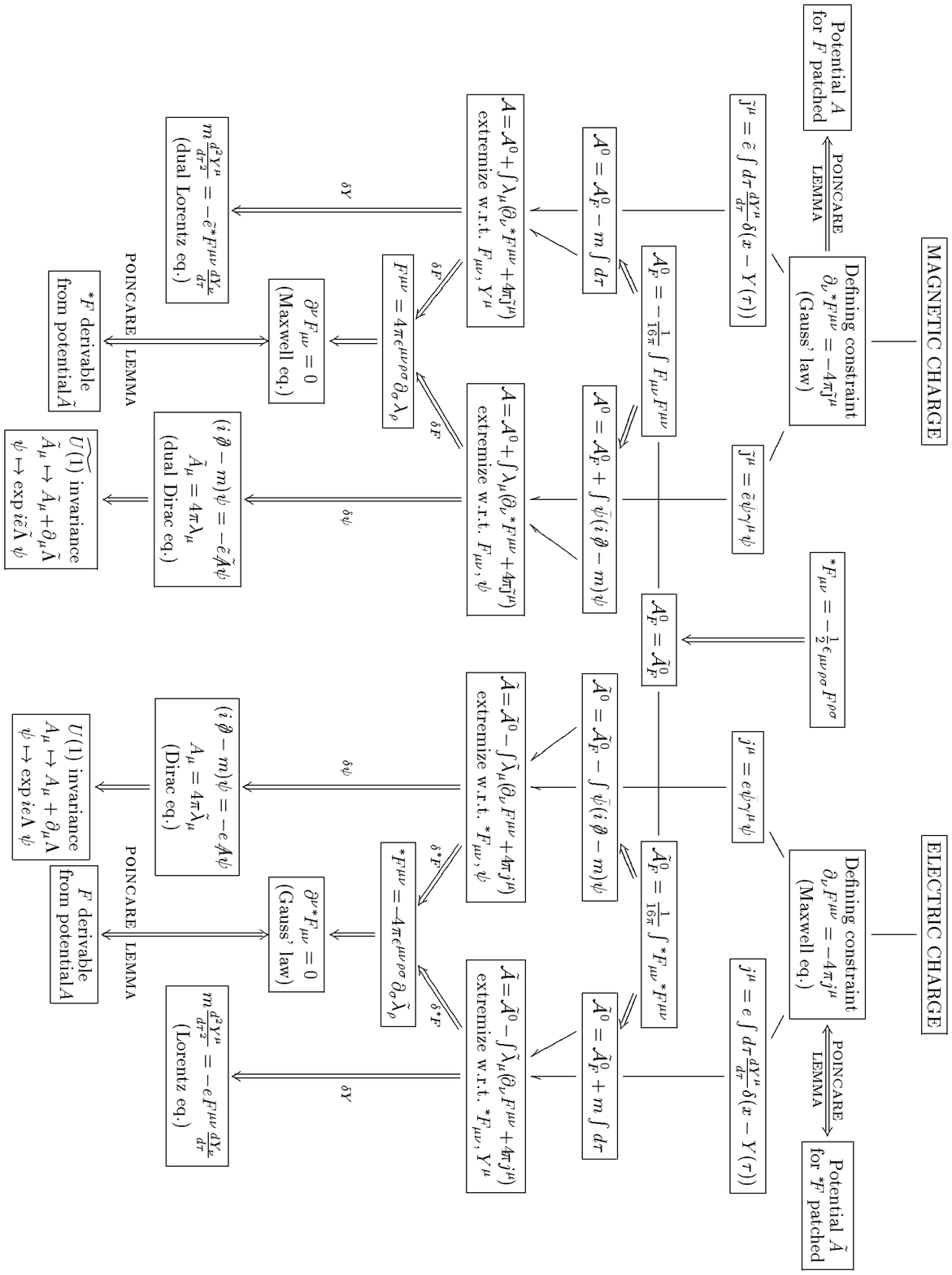}}
    }
\vspace*{-1.5cm}
 \caption[]{Electromagnetism with charges.}
  \label{chart2}
\end{chart}

\clearpage

\begin{chart}
\vspace*{-2cm}
  \centerline{
    \resizebox{!}{23cm}
{\includegraphics{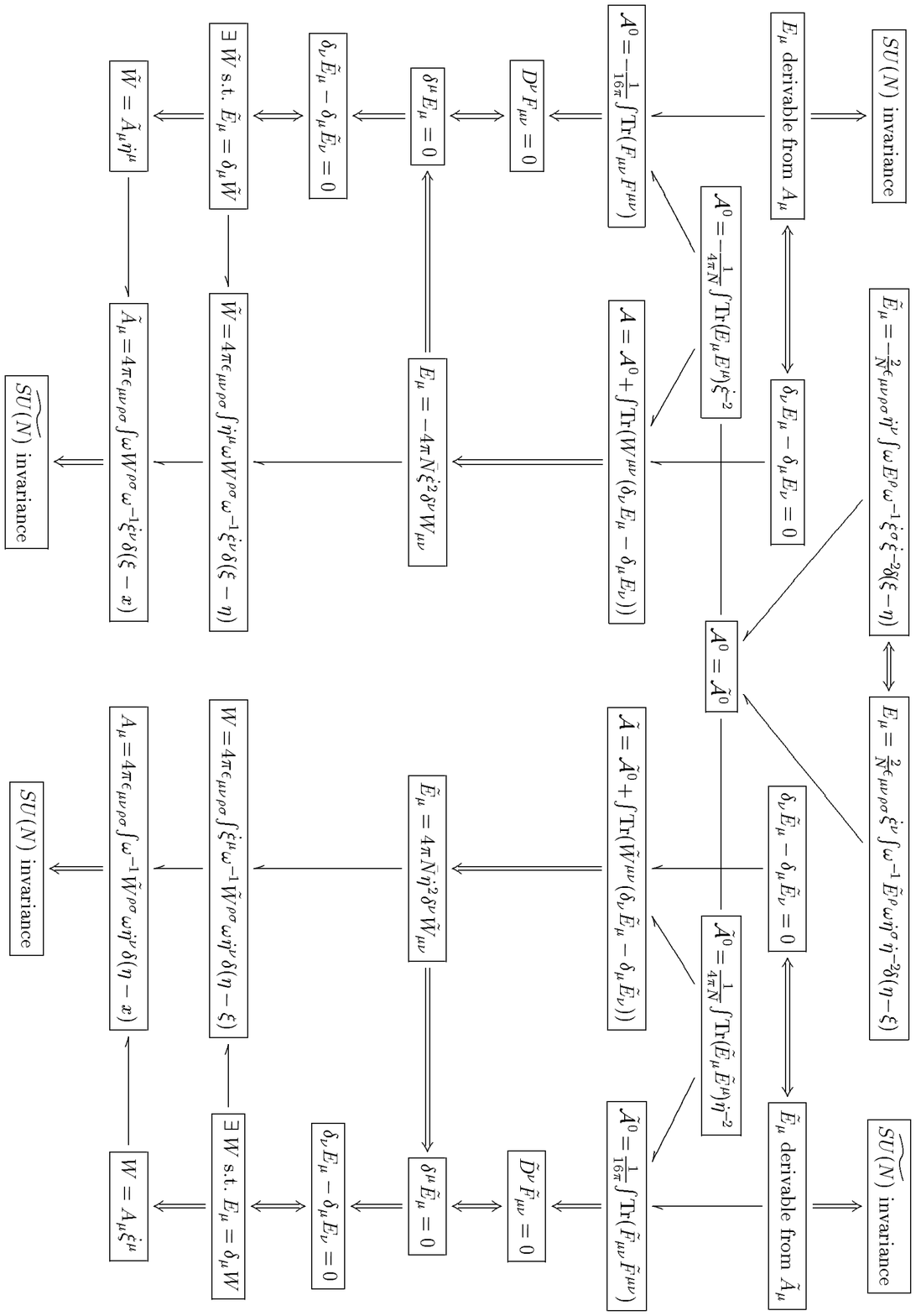}}
    }
\vspace*{-1.5cm}
 \caption[]{Pure Yang--Mills Theory.}
  \label{chart6}
\end{chart}

\clearpage

\begin{chart}
\vspace*{-2cm}
  \centerline{
    \resizebox{!}{23cm}
{\includegraphics{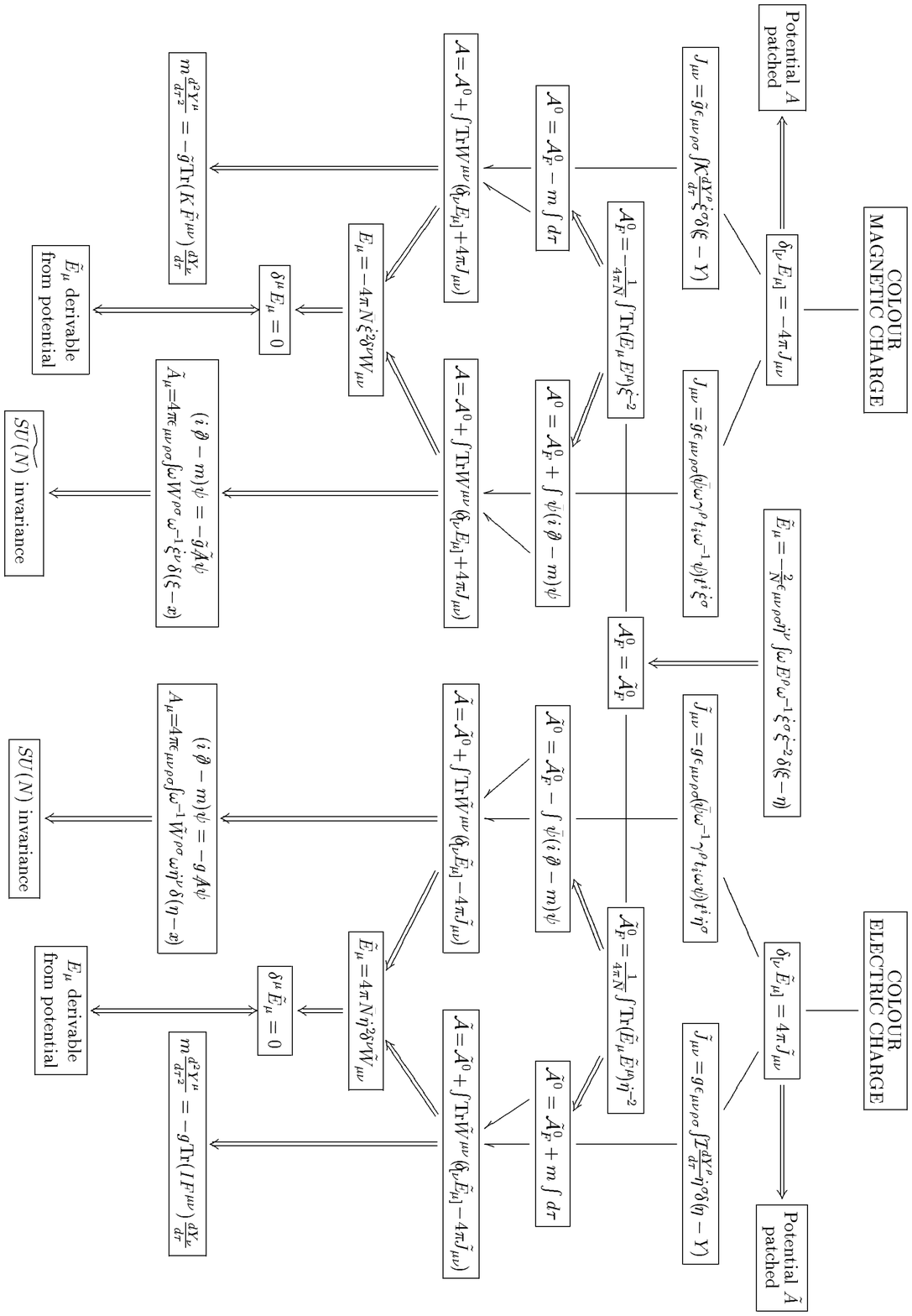}}
    }
\vspace*{-1.5cm}
 \caption[]{Yang--Mills with charges.}
  \label{chart7}
\end{chart}


\clearpage

\begin{thebibliography}{99}

\bibitem{DSMrph98} Chan Hong-Mo and Tsou Sheung Tsun, companion paper.  For
   a brief summary see Chan Hong-Mo, Jos\'e Bordes, and Tsou Sheung Tsun,
   hep-ph/9809272, talk given at ICHEP'98 in Vancouver, to appear in the
   conference proceedings.

\bibitem{Bohmaha} Y. Aharonov and D. Bohm, Phys. Rev. 115, 485 (1959).

\bibitem{Wuyang1} Tai Tsun Wu and Chen Ning Yang, Phys. Rev. D12, 3845 (1975).

\bibitem{Wuyang2} Tai Tsun Wu and Chen Ning Yang, Phys. Rev. D12, 3843 (1975).

\bibitem{Corhass} E. Corrigan and B. Hasslacher, Phys. Lett. 81B, 181, (1979).

\bibitem{loop2} Chan Hong-Mo and Tsou Sheung Tsun, Acta Phys. Pol. B17, 259,
   (1986).

\bibitem{Book} Chan Hong-Mo and Tsou Sheung Tsun, {\it Some Elementary Gauge
   Theory Concepts} (World Scientific, Singapore, 1993).

\bibitem{Polyakov} A.M. Polyakov, Nucl. Phys. 164, 171 (1980).

\bibitem{loop1} Chan Hong-Mo, Peter Scharbach and Tsou Sheung Tsun,
   Ann. Phys. (NY) 166, 396 (1986).

\bibitem{Zois} I. Zois, hep-th/9703069, to appear in
Rept. Math. Phys. (1998).

\bibitem{Lubkin} E. Lubkin, Ann. Phys. (NY) 23, 233 (1963).

\bibitem{Coleman} S. Coleman, in {\it New Phenomena in Subnuclear Physics}, 
   edited by A. Zichichi (Plenum, New York, 1976), p. 297.

\bibitem{Dirac} P.A.M. Dirac, Proc. Roy. Soc. London A133, 60 (1931).

\bibitem{Yang} C.N. Yang, Phys. Rev. D1, 2360 (1979).

\bibitem{monocharge} Chan Hong-Mo and Tsou Sheung Tsun, Nucl. Phys. B189, 364
   (1981); see also \cite{Book}.

\bibitem{monact1} Chan Hong-Mo, Peter Scharbach and Tsou Sheung Tsun, 
   Ann. Phys. (NY) 167, 454 (1986).

\bibitem{Wuyangcr} Tai Tsun Wu and Chen Ning Yang, Phys. Rev. D14, 437 (1976).

\bibitem{monact2} Chan Hong-Mo, Jacqueline Faridani, and Tsou Sheung Tsun,
   Phys. Rev. D51, 7040 (1995).

\bibitem{Guyang} C.H. Gu and C.N. Yang, Sci. Sin. 28, 483 (1975).

\bibitem{dualsymm} Chan Hong-Mo, Jacqueline Faridani, and Tsou Sheung Tsun,
   hep-th/9512173, Phys. Rev. D53, 7293 (1996).

\bibitem{Wong} S.K. Wong, Nuovo Cimento A65, 689 (1970).

\bibitem{dualsym} Chan Hong-Mo, Jacqueline Faridani, and Tsou Sheung Tsun,
   Phys. Rev. D52, 6134 (1995).

\bibitem{rudiments} Chan Hong-Mo, Jacqueline Faridani, Jakov Pfaudler,
   and Tsou Sheung Tsun, hep-th/9603188, Phys. Rev. D55, 5015 (1997).

\bibitem{thooft} G. 't~Hooft, Nucl. Phys. B138, 1 (1978); Acta Physica
   Austriaca Suppl. XXII, 531 (1980).

\bibitem{dualcomm} Chan Hong-Mo and Tsou Sheung Tsun, hep-th/9702117, 
   Phys. Rev. D56, 3646 (1997).

\bibitem{Zeni} See e.g. F. Fucito, M. Martellini, and M. Zeni, Nucl. Phys.
   B496, 259 (1997).

\end{thebibliography}
\end{document}